\documentclass[titlepage,pallis,twoside]{wpallis}
\usepackage{fancyh}
\usepackage[l]{floatflt}
\usepackage{epsfig}
\usepackage{amssymb}
\usepackage{latexsym}
\usepackage{times}
\usepackage{slashed,cite}
\usepackage{upgreek}
\numberwithin{equation}{section}

\usepackage{amsmath}
\usepackage{amsfonts}
\usepackage{amsbsy}
\usepackage{amscd}
\usepackage{bbm}

\newcommand{\bal}{\begin{align}}
\newcommand{\eal}{\end{align}}
\newcommand{\beqs}{\begin{subequations}}
\newcommand{\eeqs}{\end{subequations}}
\newcommand{\ec}{\end{center}}
\newcommand{\bec}{\begin{center}}

\newcommand{\eem}{\end{matrix}}
\newcommand{\bem}{\begin{matrix}}
\newcommand{\eeq}{\end{equation}}
\newcommand{\beq}{\begin{equation}}
\newcommand{\ba}{\begin{array}}
\newcommand{\ea}{\end{array}}
\newcommand{\bea}{\begin{eqnarray}}
\newcommand{\eea}{\end{eqnarray}}
\newcommand{\baq}{\begin{eqnarray}}
\newcommand{\eaq}{\end{eqnarray}}

\newcommand\eqs[2]{Eqs.~(\ref{#1}) and (\ref{#2})}

\newcommand\eqss[3]{Eqs.~(\ref{#1}), (\ref{#2}) and (\ref{#3})}

\newcommand{\ftn}{\footnotesize}

\newcommand{\ssz}{\scriptsize}
\newcommand{\TeV}{{\mbox{\rm TeV}}}

\newcommand{\GeV}{{\mbox{\rm GeV}}}
\newcommand{\eV}{{\mbox{\rm eV}}}

\newcommand{\sFref}[2]{Fig.~\ref{#1}-{\ftn\sf ({#2})}}
\newcommand{\sEref}[2]{Eq.~(\ref{#1}{\ftn\sf {#2}})}

\newcommand{\etal}{{\it et al.\/}}

\def\to{\rightarrow}

\def\llgm{\left\lgroup}
\def\rrgm{\right\rgroup}
\def\lf{\left(}
\def\rg{\right)}
\newcommand\vev[1]{\langle {#1} \rangle}
\newcommand{\Gr}{\ensuremath{\widetilde{G}}}
\newcommand{\Yb}{\ensuremath{Y_{B}}}
\newcommand{\Yg}{\ensuremath{Y_{3/2}}}
\newcommand{\Vhi}{\ensuremath{\widehat V_{\rm MI}}}
\newcommand{\Hhi}{\ensuremath{\widehat H_{\rm MI}}}
\newcommand{\Ohi}{\ensuremath{\Omega}}

\newcommand{\Khi}{\ensuremath{K}}
\newcommand{\Whi}{\ensuremath{W_{\rm MI}}}
\newcommand{\Vhio}{\ensuremath{\widehat V_{\rm MI0}}}

\newcommand{\mP}{\ensuremath{m_{\rm P}}}
\newcommand{\Mpq}{\ensuremath{M}}
\newcommand{\mpq}{\ensuremath{x_{M}}}

\newcommand{\Qef}{\ensuremath{\Lambda_{\rm UV}}}

\newcommand{\lm}{\ensuremath{\lambda_\mu}}

\def\openone{\leavevmode\hbox{\small1\kern-3.8pt\normalsize1}}

\newcommand{\fk}{\ensuremath{f_K}}
\newcommand{\ft}{\ensuremath{f_T}}
\newcommand{\fsh}{\ensuremath{f_{SH}}}
\newcommand{\fsn}{\ensuremath{f_{S\widetilde N^c_i}}}
\newcommand{\kx}{\ensuremath{k_S}}
\newcommand{\ksu}{\ensuremath{k_{SH_u}}}
\newcommand{\ksd}{\ensuremath{k_{SH_d}}}
\newcommand{\ksh}{\ensuremath{k_{SH}}}
\newcommand{\ksni}{\ensuremath{k_{S\widetilde N_i^c}}}
\newcommand{\ksn}{\ensuremath{k_{S\widetilde N^c}}}
\newcommand{\ck}{\ensuremath{c_{T}}}
\newcommand{\Gsm}{\ensuremath{G_{\rm SM}}}
\newcommand{\Gsn}{\ensuremath{\what{\Gamma}_{\rm \dph}}}
\newcommand{\GNsn}{\ensuremath{\what{\Gamma}_{\dph\to N_i^c}}}
\newcommand{\Ghsn}{\ensuremath{\what{\Gamma}_{\dph\to H}}}
\newcommand{\Gysn}{\ensuremath{\what{\Gamma}_{\dph\to XYZ}}}
\newcommand{\msn}{\ensuremath{\what m_{\rm \dph}}}
\newcommand{\aS}{\ensuremath{{\rm a}_S}}
\newcommand{\Ald}{\ensuremath{A_\lambda}}

\newcommand{\hd}{{\ensuremath{H_d}}}
\newcommand{\hu}{{\ensuremath{H_u}}}

\newcommand{\ns}{\ensuremath{n_{\rm s}}}
\newcommand{\as}{\ensuremath{a_{\rm s}}}
\newcommand{\As}{\ensuremath{A_{\rm s}}}
\newcommand{\rcc}{\ensuremath{\mathcal{R}}}
\newcommand{\rce}{\ensuremath{\widehat{\mathcal{R}}}}
\newcommand{\Ve}{\ensuremath{\widehat{V}}}
\newcommand{\He}{\ensuremath{{\what H}}}
\newcommand{\Ne}{\ensuremath{{\what N}}}
\newcommand{\sni}{\ensuremath{N^c_i}}
\newcommand{\ssni}{\ensuremath{\widetilde N^c_i}}

\newcommand{\mrh[1]}{\ensuremath{M_{#1N^c}}}
\newcommand{\lrh[1]}{\ensuremath{\lambda_{#1N^c}}}

\newcommand{\mD[1]}{\ensuremath{m_{#1\rm D}}}
\newcommand{\mn[1]}{\ensuremath{m_{#1\rm \nu}}}

\newcommand{\wrhn[1]}{\ensuremath{N^c_{#1}}}

\newcommand{\dphi}{\ensuremath{\what{\delta\phi}}}
\newcommand{\dph}{\ensuremath{\delta\phi}}

\newcommand{\what}{\ensuremath{\widehat}}

\def\ve{\varepsilon}
\def\aal{{\bar\alpha}}
\def\bbet{{\bar\beta}}
\def\al{{\alpha}}
\def\Aal{{\bar A}}
\def\Al{{A}}
\def\bt{{\beta}}

\def\K{{\widehat{K}}}

\def\th{{\theta}}

\def\tb{{\tan\beta}}

\newcommand{\Trh}{\ensuremath{T_{\rm rh}}}
\newcommand{\sg}{\ensuremath{\phi}}

\newcommand{\sgx}{\ensuremath{\phi_\star}}
\newcommand{\sgf}{\ensuremath{\phi_{\rm f}}}
\newcommand{\xsg}{\ensuremath{x_{\phi}}}

\newcommand{\ld}{\ensuremath{\lambda}}

\newcommand{\kp}{\ensuremath{\kappa}}
\newcommand{\se}{\ensuremath{\widehat \phi}}
\newcommand{\sex}{\ensuremath{\widehat{\phi}_\star}}

\newcommand{\geu}{\ensuremath{\widehat g}}

\newcommand{\mgr}{\ensuremath{m_{3/2}}}

\newcommand{\mg}{{\ensuremath{M_{1/2}}}}
\newcommand{\sign}{{\ensuremath{\rm sign}}}
\newcommand{\am}{\ensuremath{{\rm a}_{3/2}}}

\def\trns{transplanckian}
\def\Ka{K\"{a}hler potential}
\def\Km{K\"{a}hler manifold}
\def\Kaa{K\"{a}hler~}
\def\sub{subplanckian}
\def\sup{superpotential}

\def\FHI{nSMI~}

\newcommand{\diag}{\ensuremath{{\sf diag}}}
\newcommand{\im}{\ensuremath{{\sf Im}}}

\renewcommand{\arg}{\ensuremath{{\small\sf arg}}}
\newcommand{\tr}{{\mbox{\sf\ssz T}}}

\begin{document}

\thispagestyle{empty}
%%%%%%%%%%%%%%%

\title[]{\boldmath\Large\bfseries\scshape
Linking Starobinsky-Type Inflation \\ in no-Scale Supergravity to
MSSM}

\author{\large\bfseries\scshape C. Pallis}
\address[] {\sl Departament de F\'isica Te\`orica and IFIC,\\
Universitat de Val\`encia-CSIC, \\ E-46100 Burjassot, SPAIN \\\\
Department of Physics, University of Cyprus, \\ P.O. Box 20537,
Nicosia 1678, CYPRUS}

\begin{abstract}{{\bfseries\scshape Abstract} \\
\par A novel realization of the Starobinsky inflationary
model within a moderate extension of the Minimal Supersymmetric
Standard Model (MSSM) is presented. The proposed superpotential is
uniquely determined by applying a continuous $R$ and a
$\mathbb{Z}_2$ discrete symmetry, whereas the \Ka\ is associated
with a no-scale-type $SU(54,1)/SU(54)\times
U(1)_R\times\mathbb{Z}_2$ \Km. The inflaton is identified with a
Higgs-like modulus whose the vacuum expectation value controls the
gravitational strength. Thanks to a strong enough coupling (with a
parameter $\ck$ involved) between the inflaton and the Ricci
scalar curvature, inflation can be attained even for \sub\ values
of the inflaton with $\ck\geq76$ and the corresponding effective
theory being valid up to the Planck scale. The inflationary
observables turn out to be in agreement with the current data and
the inflaton mass is predicted to be $3\cdot10^{13}~\GeV$. At the
cost of a relatively small \sup\ coupling constant, the model
offers also a re\-solution of the $\mu$ problem of MSSM.
Supplementing MSSM  by three right-handed neutrinos we show that
spontaneously arising couplings between the inflaton and the
particle content of MSSM not only ensure a sufficiently low
reheating temperature but also support a scenario of non-thermal
leptogenesis consistently with the neutrino oscillation parameters
for gravitino heavier than about $10^4~\GeV$. }
\\ \\
{\ftn \sf Keywords: Cosmology, Supersymmetric models, Supergravity, Modified Gravity};\\
{\ftn \sf PACS codes: 98.80.Cq, 11.30.Qc, 12.60.Jv, 04.65.+e, 04.50.Kd}\\[0.2cm]
\publishedin{{\sl  J. Cosmol. Astropart. Phys. \textbf{04}, 024
(2014); J. Cosmol. Astropart. Phys. \textbf{07}, 01E (2017)}}

\end{abstract} \maketitle

%\vspace*{-.5cm}

%\thispagestyle{empty}

\setcounter{page}{1} \pagestyle{fancyplain}

\rhead[\fancyplain{}{ \bf \thepage}]{\fancyplain{}{\sl Linking
Starobinsky-Type Inflation in no-Scale SUGRA to MSSM}}
\lhead[\fancyplain{}{ \sl \leftmark}]{\fancyplain{}{\bf \thepage}}
\cfoot{}

\tableofcontents\vskip-1.3cm\noindent\rule\textwidth{.4pt}%\\
%\vspace*{.3cm}

\section{Introduction}\label{intro} %

After the announcement of the recent PLANCK results
\cite{wmap,plin}, inflation based on the potential of the
Starobinsky model \cite{R2} has gained a lot of momentum
\cite{eno5, eno7, linde, kehagias, riotto, zavalos, buch, ketov}
since it predicts \cite{R2, defelice13} a (scalar) spectral index
very close to the one favored by the fitting of the observations
by the standard power-law cosmological model with \emph{cold dark
matter} ({\ftn\sf CDM}) and a cosmological constant
($\Lambda$CDM). In particular, it has been shown that
Starobinsky-type inflation can be realized within extensions of
the \emph{Standard Model} ({\ftn\sf SM}) \cite{smR2} or
\emph{Minimal SUSY SM} ({\ftn\sf MSSM}) \cite{eno9}. However, the
realization of this type of inflation within \emph{Supergravity}
({\ftn\sf SUGRA}) is not unique. Different super- and \Ka s are
proposed \cite{linde, zavalos, eno7} which result to the same
scalar potential. Prominent, however, is the idea \cite{eno5,
eno7} of implementing this type of inflation using a \Ka, $K$,
corresponding to a $SU(N,1)/ SU(N)\times U(1)$ \Km\ inspired by
the no-scale models \cite{noscale,lahanas}. Such a symmetry fixes
beautifully the form of $K$ up to an holomorphic function $\fk$
which exclusively depends on a modulus-like field and plays the
role of a varying gravitational coupling. The stabilization of the
non-inflaton accompanying field can to be conveniently arranged by
higher order terms in $K$. In this context, a variety of models
are proposed in which inflaton can be identified with either a
matter-like \cite{eno5, eno7, eno9} or a modulus-like \cite{linde,
eno7} inflaton. The former option seems to offer a more suitable
framework \cite{eno9} for connecting the inflationary physics with
a low-energy theory, such as the MSSM endowed with right handed
neutrinos, $\sni$, since the non-inflaton modulus is involved in
the no-scale mechanism of \emph{soft SUSY breaking} ({\sf\ftn
SSB}). On the other hand, the inflationary superpotential, $\Whi$,
is arbitrarily chosen and not protected by any symmetry. Given
that, the inflaton takes \trns\ values during inflation, higher
order corrections -- e.g., by non-renormalizable terms in $\Whi$
-- with not carefully tuned coefficients may easily invalidate or
strongly affect \cite{kehagias, chaotic} the predictions of an
otherwise successful inflationary scenario.

It would be interesting, therefore, to investigate if the
shortcoming above can be avoided in the presence of a strong
enough coupling of the inflaton to gravity \cite{linde1, linde2},
as done \cite{nmi, nmN, nmH, nMCI, jones, talk} in the models of
\emph{non-minimal Inflation} ({\ftn\sf nMI}). This idea can be
implemented keeping the no-scale structure of $K$, since the
involved $\fk$ can be an analytic function, selected conveniently.
In view of the fact that $\fk$ depends only on a modulus-like
field, we here focus on this kind of inflaton -- contrary to
\cref{eno9}. As a consequence, the direct connection of the
inflationary model with the mechanism of the SSB is lost. Note, in
passing, that despite their attractive features, the no-scale
models \cite{eno9} of SSB enface difficulties -- e.g., viable SUSY
spectra are obtained only when the boundary conditions for the SSB
terms are imposed beyond the \emph{Grand Unified Theory} ({\sf\ftn
GUT}) scale and so the low energy observables depend on the
specific GUT.

Focusing on a modulus-like inflaton, the link to MSSM can be
established through the adopted $\Whi$. Its form in our work is
fixed by imposing a continuous $R$ symmetry, which reduces to the
well-known $R$-parity of MSSM, and a $\mathbb{Z}_2$ discrete
symmetry. As a consequence, $\Whi$ resembles the one used in the
widely employed models \cite{susyhybrid,dvali} of standard
\emph{F-term Hybrid Inflation} ({\sf\ftn FHI}) -- with singlet
waterfall field though. As a bonus, a dynamical generation of the
reduced Planck scale arises in \emph{Jordan Frame} ({\ftn\sf JF})
through the \emph{vacuum expectation value} ({\ftn\sf v.e.v}) of
the inflaton. Therefore the inflaton acquires a higgs-character as
in the theories of induced gravity \cite{induced, higgsflaton}. To
produce an inflationary plateau with the selected $\Whi$, $\fk$ is
to be taken quadratic, in accordance with the adopted symmetries.
This is to be contrasted with the so-called modified Cecotti model
\cite{old,linde,kehagias,zavalos,eno7} where the inflaton appears
linearly in the super- and \Ka s. The inclusion of two extra
parameters compared to the original model -- cf.
\cite{linde,kehagias,eno7} -- allows us to attain inflationary
solutions for \sub\ values of the inflaton with the successful
inflationary predictions of the model being remained intact. As a
bonus, the \emph{ultaviolet} ({\sf\ftn  UV}) cut-off scale
\cite{cutoff,cutof,riotto} of the theory can be identified with
the Planck scale and so, concerns regarding the naturalness of the
model can be safely eluded.

Our inflationary model -- let name it for short \emph{no-scale
modular inflation} ({\ftn\sf nSMI}) -- has ramifications to other
fundamental open problems of the MSSM and post-inflationary
cosmological evolution. As a consequence of the adopted $U(1)_R$
symmetry, the generation \cite{dvali,anshafi} of the mixing term
between the two electroweak Higgses is explained via the v.e.v of
the non-inflaton accompanying field, provided that a coupling
constant in $\Whi$ is rather suppressed. Finally, the observed
\cite{plcp} \emph{baryon asymmetry of the universe} ({\ftn\sf
BAU}) can be explained via spontaneous \cite{Idecay, spontaneous}
\emph{non-thermal leptogenesis} ({\ftn \sf nTL}) \cite{lept}
consistently with the $\Gr$ constraint
\cite{gravitino,brand,kohri}, the data \cite{valle,lisi} on the
neutrino oscillation parameters as long as the masses of the
gravitino ($\Gr$) lie in the multi-$\TeV$ region -- as dictated in
many versions \cite{anomaly13,focus,pure} of MSSM after the recent
LHC \cite{lhc,mssm} results on the Higgs boson mass.

The basic ingredients -- particle content and structure of the
super- and  \Ka s -- of our model are presented in
Sec.~\ref{fhim}. In Sec.~\ref{fhi} we describe the inflationary
potential, derive the inflationary observables and confront them
with observations. Sec.~\ref{secmu} is devoted to the resolution
of the $\mu$ problem of MSSM. In Sec.~\ref{pfhi} we outline the
scenario of nTL, exhibit the relevant imposed constraints and
describe our predictions for neutrino masses. Our conclusions are
summarized in Sec.~\ref{con}. Throughout the text, the subscript
of type $,\chi$ denotes derivation  \emph{with respect to}
({\ftn\sf w.r.t}) the field $\chi$ (e.g.,
$_{,\chi\chi}=\partial^2/\partial\chi^2$) and charge conjugation
is denoted by a star.

\section{Model Description}\label{fhim}

We focus on a moderated extension of MSSM with three $\sni$'s
augmented by two superfields, a matter-like $S$ and a modulus-like
$T$, which are singlets under $\Gsm=SU(3)_{\rm c}\times SU(2)_{\rm
L}\times U(1)_{Y}$. Besides the local symmetry of MSSM, $\Gsm$,
the model possesses also the baryon number symmetry $U(1)_B$, a
nonanomalous $R$ symmetry $U(1)_{R}$ and a discrete
$\mathbb{Z}_2$. Note that global continuous symmetries can
effectively arise \cite{laz1} in many compactified string
theories. The charge assignments under the global symmetries of
the various matter and Higgs superfields are listed in
Table~\ref{tab1}.  We below present the structure of the
superpotential (\Sref{sup}) and the \Ka\ (\Sref{ka}) of our model.

\subsection{The Superpotential} \label{sup}

The superpotential of our model naturally splits into two parts:
\beqs\beq W=W_{\rm MSSM}+W_{\rm MI}, \label{Wol}\eeq
where $W_{\rm MSSM}$ is the part of $W$ which contains the usual
terms -- except for the $\mu$ term -- of MSSM, supplemented by
Yukawa interactions among the left-handed leptons and $\sni$
\beq W_{\rm MSSM} = h_{ijE} {e}^c_i {L}_j \hd+h_{ijD} {d}^c_i
{Q}_j \hd + h_{ijU} {u}^c_i {Q}_j \hu+ h_{ijN} \sni L_j \hu.
\label{wmssm}\eeq
Here the $i$th generation $SU(2)_{\rm L}$ doublet left-handed
quark and lepton superfields are denoted by $Q_i$ and $L_i$
respectively, whereas the $SU(2)_{\rm L}$ singlet antiquark
[antilepton] superfields by $u^c_i$ and ${d_i}^c$ [$e^c_i$ and
$\sni$] respectively. The electroweak Higgs superfields which
couple to the up [down] quark superfields are denoted by $\hu$
[$\hd$].

On the other hand, $W_{\rm MI}$ is the part of $W$ which is
relevant for nSMI, the generation of the $\mu$ term of MSSM and
the Majorana masses for $\sni$'s. It takes the form
\beq\label{Whi} W_{\rm MI}= \ld S\lf T^2-\Mpq^2/2\rg+\lm
S\hu\hd+\frac12\mrh[i] N^{c2}_i+\lrh[ij]T^2N^{c}_iN^{c}_j/2\mP,
\eeq\eeqs
where $\mP = 2.44\cdot 10^{18}~\GeV$ is the reduced Planck mass.
The imposed $U(1)_R$ symmetry ensures the linearity of $\Whi$
w.r.t $S$. This fact allows us to isolate easily via its
derivative the contribution of the inflaton $T$ into the F-term
SUGRA scalar potential, placing $S$ at the origin. The imposed
$\mathbb{Z}_2$ prohibits the existence of the term $ST$ which,
although does not drastically modifies our proposal, it
complicates the determination of SUSY vacuum and the inflationary
dynamics. On the other hand, the imposed symmetries do not forbid
non-renormalizable terms of the form $T^{2n+2}$ where $n\geq1$ is
an integer. For this reason we are obliged to restrict ourselves
to \sub\ values of $T$.

The second term in the \emph{right-hand side} ({\sf\ftn r.h.s}) of
\Eref{Whi} provides the $\mu$ term of MSSM along the lines of
Ref.~\cite{dvali,anshafi} -- see \Sref{secmu}. The third term is
the Majorana mass term for the $\sni$'s and we assume that it
overshadows (for sufficiently low $\lrh[ij]$'s) the last
non-renormalizable term which is neglected henceforth. Here we
work in the so-called \emph{right-handed neutrino basis}, where
$\mrh[i]$ is diagonal, real and positive. These masses together
with the Dirac neutrino masses in Eq.~(\ref{wmssm}) lead to the
light neutrino masses via the seesaw mechanism. The same term is
important for the decay \cite{Idecay,spontaneous} of the inflaton
after the end of \FHI to $\ssni$, whose subsequent decay can
activate nTL. As a result of the imposed $\mathbb{Z}_2$, a term of
the form $TN_i^{c2}$ is prohibited and so the decay of $T$ into
$\sni$ is processed by suppressed SUGRA-induced interactions
\cite{Idecay}, guaranteing thereby a sufficiently low reheat
temperature compatible with the $\Gr$ constraint and successful
nTL -- see \Sref{lept}.

\begin{table}[!t]\bec
\begin{tabular}{|c|cccccccccc|}\hline
& $S$ & $T$& $\hu$ & $\hd$ & $L_i$ & $\sni$ & $e^c_i$ & $Q_i$ &
$u^c_i$ & $d^c_i$ \\ \hline $U(1)_B$ & 0 & 0 & 0 &0 & 0 & 0 & 0 &
1/3 & -1/3 & -1/3\\ $U(1)_R$ & 2 & 0 & 0 &0 & 1 & 1 & 1 & 1 & 1
& 1\\ $\mathbb{Z}_2$ & 0 & 1 & 0 & 0 & 0 & 0 & 0 & 0 & 0&0\\
\hline
\end{tabular}\ec
\hfill \vchcaption[]{\sl\small  The global charges of the
superfields of our model.}\label{tab1}
\end{table}

Since the no-scale SUGRA adopted here leads to the
non-renormalizable F-term (scalar) potential in \Eref{Vsugra}, we
expect that it yields an effective SUSY theory which depends not
only on the superpotential, $W_{\rm MI}$ in \Eref{Whi}, but also
on the \Ka, $K$ in \Eref{Kol} -- see \Sref{ka}. To trace out this
behavior we apply the generic formula for the SUSY F-term
potential \cite{martin}:
%-- cf. \srEref{2.2}{a}
\beq \label{Vsusy} V_{\rm SUSY}= K^{\al\bbet} W_{\rm MI\al}
W^*_{\rm MI\bbet}\,,\eeq
which is obtained from the SUGRA potential in \Eref{Vsugra} -- see
\Sref{fhi1} below -- if we perform an expansion in powers $1/\mP$
and take the limit $\mP\to\infty$. The \Ka, $K$, employed here can
not be expanded in powers of $1/\mP$, since unity is not included
in the argument of the logarithm -- in contrast to the $K$'s used
in \cref{jhep,var}. In \Eref{Vsusy} $K^{\al\bbet}$ is the inverse
of the K\"ahler metric $K_{\al\bbet}$ with $z^\al=T, S, \hu, \hd$
and $\ssni$  where the complex scalar components of the
superfields $T, S, \hu$ and $\hd$ are denoted by the same symbol
whereas this of $\sni$ by $\ssni$. We find that $K_{\al\bbet}$
reads
\beq \lf K_{\al\bbet}\rg=\lf{3\over\Ohi}\rg^2\llgm\bem
{12\ck^2|T|^2\over\mP^2} & -{2\ck ST\over\mP^2}&-{2\ck \hu
T\over\mP^2}&-{2\ck \hd T\over\mP^2}&-{2\ck \ssni T\over\mP^2}\cr
-{2\ck
S^*T^*\over\mP^2}&-{\Ohi\over3}&{S^*H_u\over3\mP^2}&{S^*H_d\over3\mP^2}&{S^*\ssni\over3\mP^2}\cr
-{2\ck
H_u^*T^*\over\mP^2}&{H_u^*S\over3\mP^2}&-{\Ohi\over3}&{H_u^*\hd\over3\mP^2}&{H_u^*\ssni\over3\mP^2}\cr
-{2\ck
H_d^*T^*\over\mP^2}&{H_d^*S\over3\mP^2}&{H_d^*\hu\over3\mP^2}&-{\Ohi\over3}&{H_d^*\ssni\over3\mP^2}\cr
-{2\ck \widetilde N_i^{c*}T^*\over\mP^2}&{\widetilde
N_i^{c*}S\over3\mP^2}&{\widetilde N_i^{c*}\hu\over3\mP^2}&{\ck
\widetilde N_i^{c*}\hd\over3\mP^2}&-{\Ohi\over3}\eem\rrgm\,,\eeq
where $\Ohi$ is given in \Eref{minK} and we neglect the fourth
order terms since we expect that these are not relevant for the
low energy effective theory. The inverse of the matrix above is
\beq \lf K^{\al\bbet}\rg=-{\Ohi\over3}\llgm\bem
(T^2+T^{*2})/12\ck|T|^2 & S/6\ck T^*&\hu/6\ck T^*&\hd/6\ck
T^*&\ssni/6\ck T^* \cr
S^*/6\ck T&1&0&0&0\cr
H^*_u/6\ck T&0&1&0&0\cr
H_d^*/6\ck T&0&0&1&0\cr
\widetilde N_i^{c*}/6\ck T&0&0&0&1\eem\rrgm\,.\eeq
Substituting this in \Eref{Vsusy}, we end up with the following
expression
\beqs\bea \nonumber  V_{\rm SUSY}
&=&-{\Ohi\over3}\Bigg(\ld^2\left|T^2 + \lm \hu\hd/\ld
-M^2/2\right|^2+\lm^2\lf|\hu|^2 + |\hd|^2 \rg|S|^2
+\mrh[i]^2|\widetilde N^{c}_i|^2\\
\nonumber &&+\frac{2\ld^2}{3\ck}|S|^2\lf
T^2+T^{*2}-M^2/2\rg+\frac{\ld
\lm}{\ck} \lf \hu\hd +H^*_uH^*_d\rg|S|^2\\
&&+\frac{\ld}{3\ck}\mrh[i]\lf S^* \widetilde N_i^{c2}+S \widetilde
N_i^{c*2}\rg\Bigg)\,. \label{VF}\eea
The three first terms in the r.h.s of the expression above come
from the terms $K^{\al\aal} W_{\rm MI\al} W^*_{\rm MI\aal}$ of
\Eref{Vsusy} for $z^\al=S,\hu,\hd,\ssni$. The fourth one comes
from the terms $$K^{\al\aal} W_{\rm MI\al} W^*_{\rm
MI\aal}+K^{\al\bbet} W_{\rm MI\al} W^*_{\rm MI\bbet}+K^{\bt\aal}
W_{\rm MI\bt} W^*_{\rm MI\aal}$$ for $z^\al=T$ and $z^\bt=S$; the
residual terms arise from terms of the form $K^{\al\bbet} W_{\rm
MI\al} W^*_{\rm MI\bbet}+K^{\bt\aal} W_{\rm MI\bt} W^*_{\rm
MI\aal}$; for $z^\al=T$ and $z^\bt=\hu, \hd$ the fifth one and
$z^\bt=\ssni$ the last one.

From the potential in Eq.~(\ref{VF}), we find that the SUSY vacuum
lies at
\beq \vev{\hu}=\vev{\hd}=\vev{\ssni}=0,~\vev{S}\simeq0
\>\>\mbox{and}\>\> \sqrt{2}|\vev{T}|=M.\label{vevs} \eeq\eeqs
Contrary to the Cecotti model \cite{old,linde,eno7} our modulus
$T$ can take values $M\leq\mP$ at the SUSY vacuum. Also, $\vev{T}$
breaks spontaneously the imposed $\mathbb{Z}_2$ and so, it can
comfortably decay via SUGRA-inspired decay channels -- see
\Sref{lept} -- reheating the universe and rendering
\cite{spontaneous} spontaneous nTL possible. No domain walls are
produced due to the spontaneous breaking of $\mathbb{Z}_2$ at the
SUSY vacuum, since this is broken already during nSMI.

With the addition of SSB terms, as required in a realistic model,
the position of the vacuum shifts \cite{dvali,anshafi} to non-zero
$\vev{S}$ and an effective $\mu$ term is generated from the second
term in the r.h.s of \Eref{Whi} -- see \Sref{secmu}. Let us
emphasize that SSB effects explicitly break $U(1)_R$ to the
$\mathbb{Z}_2^{R}$ matter parity, under which all the matter
(quark and lepton) superfields change sign. Combining
$\mathbb{Z}_2^{R}$ with the $\mathbb{Z}_2^{\rm f}$ fermion parity,
under which all fermions change sign, yields the well-known
$R$-parity. Recall that this residual symmetry prevents the rapid
proton decay, guarantees the stability of the \emph{lightest SUSY
particle} ({\sf \ftn LSP}) and therefore it provides a
well-motivated CDM candidate. Needless to say, finally, that such
a connection of the Starobinsky-type inflation with this vital for
MSSM $R$-symmetry can not be established within the modified
Cecotti model \cite{old,linde,zavalos}, since no symmetry can
prohibit a quadratic term for the modulus-like field in
conjunction with the tadpole term in $\Whi$.

\subsection{The \Kaa Potential} \label{ka}

According to the general discussion of \cref{noscale}, the \Km\
which corresponds to a \Ka\ of the form
%
%\beq \Khi=-3\mP^2\ln\lf {\ck\over\mP^2}\lf
%T^2+T^{*2}\rg-{|X|^2\over3\mP^2}+{\kx}{|S|^4\over\mP^4}+\cdots\rg,\label{minK}\eeq
%
\beqs\beq \Khi=-3\mP^2\ln\lf
\fk(T)+\fk^*(T^*)-{\Phi_\Al\Phi^{*\Aal}\over3\mP^2}+k_{S\Phi^\Al}{|S|^2|\Phi_\Al|^2\over3\mP^4}+\cdots\rg,\label{Kol}\eeq
with $\fk$ being an holomorphic function of $T$, exhibits a $SU(N,
1)/SU(N) \times U(1)_R\times \mathbb{Z}_2$ global symmetry. Here
$N-1=53$ is the number of scalar components of $S,~\sni$ and the
MSSM superfields which are collectively denoted as \beq
\Phi^\Al=\tilde e_i^c,~\tilde u_i^c,~\tilde d_i^c,~\ssni,~\tilde
L_i,~\widetilde Q_i,~\hu,~\hd \>\>\mbox{and}\>\>S. \label{fas}\eeq
Note that summation over the repeated (small or capital) Greek
indices is implied. The third term in the r.h.s of \Eref{Kol} --
with coefficients $k_{S\Phi^\Al}$ being taken, for simplicity,
real -- is included since it has an impact on the scalar mass
spectrum along the inflationary track -- see \Sref{fhi1}. In
particular, the term with coefficient $k_{SS}=\kx\simeq1$ assists
us to avoid the tachyonic instabilities encountered in similar
models \cite{linde, eno7, zavalos, linde1, linde2} -- see
\Sref{fhi1}. The ellipsis represents higher order terms which are
irrelevant for the inflationary dynamics since they do not mix the
inflaton $T$ with the matter fields. This is, in practice, a great
simplification compared to the models of nMI -- cf.~\cref{talk}.
Contrary to other realizations of the Starobinsky model -- cf.
\cref{linde, eno7,zavalos} --, we choose $\fk$ to be quadratic and
not linear with respect to $T$, i.e., \beq \fk(T)=\ck T^2/\mP^2
\label{fdef}\eeq\eeqs in accordance with the imposed
$\mathbb{Z}_2$ symmetry which forbids a linear term -- the
coefficient $\ck$ is taken real too. As in the case of \Eref{Whi},
non-renormalizable terms of the form $T^{2n+2}$, with integer
$n\geq1$, are allowed but we can safely ignore them restricting
ourselves to $T\leq\mP$.

%holomorphic part
The interpretation of the adopted $K$ in \Eref{Kol} can be given
in the ``physical'' frame by writing the JF action for the scalar
fields $\Phi^\al=\Phi^A,T$. To extract it, we start with the
corresponding EF action within SUGRA \cite{linde1,nmN,talk} which
can be written as
\beqs \beq\label{Saction1} {\sf S}=\int d^4x \sqrt{-\what{
\mathfrak{g}}}\lf-\frac{1}{2}\mP^2 \rce +K_{\al\bbet} \dot\Phi^\al
\dot\Phi^{*\bbet}-\Ve_{\rm MI0}+\cdots\rg,\eeq
where $K_{\al\bbet}={\K_{,\Phi^\al\Phi^{*\bbet}}}$ with
$K^{\bbet\al}K_{\al\bar \gamma}=\delta^\bbet_{\bar \gamma}$,
$\widehat{\mathfrak{g}}$ is the determinant of the EF metric
$\geu_{\mu\nu}$, $\rce$ is the EF Ricci scalar curvature,
$\Ve_{\rm MI0}$ is defined in \Sref{fhi1}, the dot denotes
derivation w.r.t the JF cosmic time and the ellipsis represents
terms irrelevant for our analysis. Performing then a suitable
conformal transformation, along the lines of \cref{nmN, talk} we
end up with the following action in the JF
\beq {\sf S}=\int d^4x
\sqrt{-\mathfrak{g}}\lf-\frac{\mP^2}{2}\lf-{\Omega\over3}\rg
\rcc+\mP^2\Omega_{\al{\bbet}}\dot \Phi^\al \dot
\Phi^{*\bbet}-V_{\rm SUSY}+\cdots\rg , \label{Sfinal}\eeq
where $g_{\mu\nu}=-\lf 3/\Omega\rg\geu_{\mu\nu}$ is the JF metric
with determinant $\mathfrak{g}$, $\rcc$ is the JF Ricci scalar
curvature, and we use the shorthand notation
$\Omega_\al=\Omega_{,\Phi^\al}$ and
$\Omega_\aal=\Omega_{,\Phi^{*\aal}}$. The corresponding frame
function can be found from the relation
%
%\beq -{\Ohi\over3}=e^{-\Khi/3\mP^2}= {\ck\over\mP^2}\lf
%T^2+T^{*2}\rg-{\Phi_a\Phi^{*\aal}\over3\mP^2}+{\kx}{|S|^4\over\mP^4}+\cdots,
%\label{minK}\eeq
\beq -{\Ohi\over3}=e^{-\Khi/3\mP^2}=
\fk(T)+\fk^*(T^*)-{\Phi_A\Phi^{*\Aal}\over3\mP^2}+k_{S\Phi^\Al}{|S|^2|\Phi_\Al|^2\over3\mP^4}+\cdots\,\cdot
\label{minK}\eeq\eeqs
The last result reveals that $T$ has no kinetic term, since
$\Omega_{,TT^*}=0$. This is a crucial difference between the
Starobinsky-type models and those \cite{talk} of nMI, with
interest consequences \cite{riotto} to the derivation of the
ultraviolet cutoff scale of the theory -- see \Sref{fhi2}.
Furthermore, given that $\vev{\Phi^A}\simeq0$, recovering the
conventional Einstein gravity at the SUSY vacuum, \Eref{vevs},
dictates
%
%\beq \ck\lf \fk(\vev{T}^2+\vev{T^*}^2\rg=\mP^2 \label{vev}\eeq
\beq
\fk(\vev{T})+\fk^*(\vev{T^*})=1~~\Rightarrow~~M=\mP/\sqrt{\ck}.
\label{ig}\eeq Given that the analysis of inflation in both frames
yields equivalent results \cite{defelice13, defelice}, we below --
see Sec.~\ref{fhi1} and \ref{fhi2} -- carry out the derivation of
the inflationary observables exclusively in the EF.

\section{The Inflationary Scenario}\label{fhi}

In this section we outline the salient features of our
inflationary scenario (\Sref{fhi1}) and then, we present its
predictions in Sec.~\ref{num1}, calculating a number of observable
quantities introduced in Sec.~\ref{fhi2}. We also provide a
detailed analysis of the UV behavior of the model in
Sec.~\ref{fhi3}.

\subsection{The Inflationary
Potential}\label{fhi1}

The EF F--term (tree level) SUGRA scalar potential, $\Vhio$, of
our model -- see \Eref{Saction1} -- is obtained from $\Whi$ in
Eq.~(\ref{Whi}) and $\Khi$ in \Eref{Kol} by applying the standard
formula:
\beq \Vhio=e^{\Khi/\mP^2}\left(K^{\al\bbet}{\rm F}_\al {\rm
F}^*_\bbet-3\frac{\vert W_{\rm
MI}\vert^2}{\mP^2}\right),\label{Vsugra} \eeq where ${\rm
F}_\al=W_{\rm MI,\Phi^\al} +K_{,\Phi^\al}W_{\rm MI}/\mP^2$.
Setting the fields $\Phi^\al=S,\ssni,\hu$ and $\hd$ at the origin
the only surviving term of $\Vhio$ is
\beq \Vhio = e^{K/\mP^2}K^{SS^*}\, W_{{\rm MI},S}\, W^*_{{\rm
MI},S^*}= \frac{\ld^2 |2T^2-M^2|^2}{4
(\fk+\fk^*)^2}\cdot\label{Vhi0}\eeq
It is obvious from the result above that a form of $\fk$ as the
one proposed in \Eref{fdef} can flatten $\Vhio$ sufficiently so
that it can drive nSMI.  Employing the dimensionless variables
\beq \label{dimlss}
\xsg=\sg/\mP,\>\>\>\ft=1-\ck\xsg^2\>\>\>\mbox{and}\>\>\>\mpq=M/\mP\>\>\>\mbox{with}\>\>\>\sg=|T|/\sqrt{2}
\eeq
and setting $\arg T=0$, $\Vhio$ and the corresponding Hubble
parameter $\He_{\rm MI}$ read
\beq \Vhio=\frac{\ld^2 \mP^4(\xsg^2-\mpq^2)^2}{4
\ck^2\xsg^4}=\frac{\ld^2 \mP^4 \ft^2}{4 \ck^4 \xsg^4}
\>\>\>\mbox{and}\>\>\> \He_{\rm
MI}={\Vhio^{1/2}\over\sqrt{3}\mP}\simeq{\ld\mP\over2\sqrt{3}\ck^2}\,,\label{Vhio}\eeq
where we put $\mpq=1/\sqrt{\ck}$ -- by virtue of \Eref{ig} -- in
the final expressions.

Expanding $T$ and $\Phi^\al$ in real and imaginary parts as
follows
\beq T= \frac{\phi}{\sqrt{2}}\,e^{i
\th/\mP}\>\>\>\mbox{and}\>\>\>X^\al= \frac{x^\al +i\bar
x^\al}{\sqrt{2}}\>\>\>\mbox{with}\>\>\>X^\al=S,\hu,\hd,\ssni\label{cannor}
\eeq
we can check the stability of the inflationary direction
\beq \th=x^\al=\bar
x^\al=0\>\>\mbox{where}\>\>x^\al=s,h_u,h_d,\tilde\nu^c_i,\label{inftr}
\eeq
w.r.t the fluctuations of the various fields. In particular, we
examine the validity of the extremum and minimum conditions, i.e.,
\beqs\beq \left.{\partial
\Vhio\over\partial\what\chi^\al}\right|_{\mbox{\Eref{inftr}}}=0\>\>\>
\mbox{and}\>\>\>\what m^2_{
\chi^\al}>0\>\>\>\mbox{with}\>\>\>\chi^\al=\th,x^\al,\bar
x^\al.\label{Vcon} \eeq
Here $\what m^2_{\chi^\al}$ are the eigenvalues of the mass matrix
with elements
\beq \label{wM2}\what
M^2_{\al\bt}=\left.{\partial^2\Vhio\over\partial\what\chi^\al\partial\what\chi^\beta}\right|_{\mbox{\Eref{inftr}}}
\mbox{with}~~\chi^\al=\th,x^\al,\bar x^\al\eeq\eeqs
and hat denotes the EF canonically normalized fields. Taking into
account that along the configuration of \Eref{inftr}
$K_{\al\bbet}$ defined below \Eref{Saction1} takes the form
%\succapprox
\beq \lf K_{\al\bbet}\rg=\diag\lf
6/\xsg^2,\underbrace{1/\ck\xsg^2,...,1/\ck\xsg^2}_{8~\mbox{\ftn
elements}}\rg \label{VJe3}\eeq -- here we take into account that
$\hu$ and $\hd$ are $SU(2)_{\rm L}$ doublets --, the kinetic terms
of the various scalars in \Eref{Saction1} can be brought into the
following form
\beqs\beq \label{K3} K_{\al\bbet}\dot\Phi^\al
\dot\Phi^{*\bbet}=\frac12\lf\dot{\se}^{2}+\dot{\what
\th}^{2}\rg+\frac12\lf\dot{\what x}_\al\dot{\what x}^\al
+\dot{\what{\overline x}}_\al\dot{\what{\overline x}}^\al\rg,\eeq
where the hatted fields are defined as follows
\beq  \label{cannor3b} {d\widehat \sg/d\sg}=J=\sqrt{6}/\xsg,\>\>\>
\what{\th}= \sqrt{6}\th, \>\>\>\what x^\al=
x^\al/\sqrt{\ck}\xsg\>\>\>\mbox{and}\>\>\>\what{\bar x}^\al= \bar
x^\al/\sqrt{\ck}\xsg.\eeq\eeqs

\renewcommand{\arraystretch}{1.4}
\begin{table}[!t]
\bec\begin{tabular}{|c|c|l|}\hline
{\sc Fields} &{\sc Eingestates} & \hspace*{3.cm}{\sc Masses Squared}\\
\hline \hline
$1$ real scalar &$\what{\th}$ & $\what m^2_{\th}=
\ld^2\mP^2(\ft+2\ck^2\xsg^2)/3\ck^4\xsg^4\simeq4\He_{\rm MI}^2$\\
$2$ real scalars &$\what{s},~\what{\bar s}$ & $\what m^2_{
s}=\ld^2\mP^2 (1 + \ck\xsg^2 (2 - \ck\xsg^2 + 6\kx\ft^2))/6\ck^4\xsg^4$\\
$4$ real scalars &$\what{h}_+,\what{\bar h}_+$ & $\what m^2_{
h+}=\ld\mP^2\ft(\ld\ft\fsh+6 \ld_\mu \ck^2\xsg^2)/12\ck^4\xsg^4$\\
$4$ real scalars &$\what{h}_-,\what{\bar h}_-$ & $\what m^2_{
h-}=\ld\mP^2\ft(\ld\ft\fsh-6 \ld_\mu \ck^2\xsg^2)/12\ck^4\xsg^4$\\
$6$ real scalars &$\what{\tilde \nu}^c_i,\what{\bar {\tilde
\nu}}^c_i$ & $\what m^2_{i \nu^c}= (\ld^2
\mP^2\ft^2\fsn+12\mrh[i]^2\ck^3 \xsg^2)/12\ck^4 \xsg^4$\\\hline
$2$ Weyl spinors & $\what{\psi}_\pm$& $\what m^2_{
\psi\pm}\simeq\ld^2\mP^2/3\ck^4\xsg^4$\\
$3$ Weyl spinors &$\what{N}_i^c$ & $\what m^2_{iN^c}= \mrh[i]^2
/\ck\xsg^2$ \\ \hline
\end{tabular}\ec
\hfill \vchcaption[]{\sl\small The mass spectrum of our model
along the inflationary trajectory of \Eref{inftr}. }\label{tab3}
\end{table}

Upon diagonalization of the relevant sub-matrices of $\what
M^2_{\al\bt}$, \Eref{wM2}, we construct the scalar mass spectrum
of the theory along the direction in \Eref{inftr}. Our results are
summarized in \Tref{tab3}, assuming $\ksu\simeq\ksd=\ksh$ in order
to avoid very lengthy formulas for the masses of $\what h_\pm$ and
$\what{\bar h}_\pm$. The various unspecified there eigenvalues are
defined as follows:
\beqs\beq \what h_\pm=(\what h_u\pm{\what h_d})/\sqrt{2},\>\>\>
\what{\bar h}_\pm=(\what{\bar h}_u\pm\what{\bar
h}_d)/\sqrt{2}\>\>\>\mbox{and}\>\>\>\what
\psi_\pm={(\what{\psi}_{T}\pm \what{\psi}_{S})/\sqrt{2}}, \eeq
where the spinors $\psi_T, \psi_S$ and $N^c_i$ associated with the
superfields $S, T$ and $N^c_i$ are related to the normalized ones
in \Tref{tab3} as follows:
\beq \label{psis} \what\psi_{S}=\sqrt{6}\psi_{S}/
\xsg,\>\>\>\what\psi_{T}=\psi_{T}/\sqrt{\ck}\xsg\>\>\>
\mbox{and}\>\>\>\what{N}_i^c=N_i^c/\sqrt{\ck}\xsg. \eeq\eeqs
We also use the shorthand notation:
\beq \label{fss}
\fsh=2+3\ksh\ck\xsg^2\>\>\>\mbox{and}\>\>\>\fsn=2+3\ksni\ck\xsg^2.\eeq
Note that, due to the large effective masses that the $\chi$'s in
\Eref{wM2} acquire during nSMI, they enter a phase of oscillations
about $\chi=0$ with reducing amplitude. As a consequence -- see
\Eref{cannor3b} --, $\dot{\widehat \chi}\simeq\dot
\chi/\sqrt{\fk}$ since the quantity $\dot \fk/2\fk^{3/2}\chi$,
involved in relating $\dot{\widehat \chi}$ to $\dot \chi$, turns
out to be negligibly small compared with $\dot{\chi}/\sqrt{\fk}$
-- cf.~\cref{nMCI}. Moreover, we have numerically verified that
the various masses remain greater than $\Hhi$ during the last $50$
e-foldings of nSMI, and so any inflationary perturbations of the
fields other than the inflaton are safely eliminated -- see also
\Sref{num1}.

From \Tref{tab3} it is evident that $\kx\gtrsim1$ assists us to
achieve $\what m^2_{{s}}>0$ -- in accordance with the results of
\cref{linde,eno7,zavalos}. On the other hand, given that
$\ft\leq0$, $\what m^2_{h-}>0$ requires
\beq \label{lm}\ld\ft\fsh+6\ld_\mu
\ck^2\xsg^2<0~~\Rightarrow~~\lm<-\frac{\ld\ft\fsh}{6\ck^2\xsg^2}
\simeq{\ld\over3\ck}+\frac12{\ld\ksh\xsg^2}\simeq2\cdot10^{-5}-10^{-6},
\eeq
as $\ksh$ decreases from $3$ to $0.5$. Here we have made use of
\eqs{sgap}{lan} -- see \Sref{fhi2}. We do not consider such a
condition on $\lm$ as unnatural, given that $h_{11U}$ in
\Eref{wmssm} is of the same order of magnitude too -- cf.
\cref{fermionM}. In \Tref{tab3} we also present the masses squared
of chiral fermions along the trajectory of \Eref{inftr}, which can
be served for the calculation of the one-loop radiative
corrections. Employing the well-known Coleman-Weinberg formula
\cite{cw}, we find that the one-loop corrected inflationary
potential is
\bea\Vhi=\Vhio&+&{1\over64\pi^2}\lf \widehat m_{
\th}^4\ln{\widehat m_{\th}^2\over\Lambda^2} +2 \widehat m_{
s}^4\ln{\widehat m_{s}^2\over\Lambda^2} + 4\widehat m_{
h+}^4\ln{\widehat m_{h+}^2\over\Lambda^2}+ 4\widehat m_{
h-}^4\ln{\widehat m_{h-}^2\over\Lambda^2}
\nonumber \right.\\
&+&\left.2\sum_{i=1}^3\lf\widehat m_{ i\nu^c}^4\ln{\widehat
m_{i\nu^c}^2\over\Lambda^2}-\widehat m_{ iN^c}^4\ln{\widehat
m_{iN^c}^2\over\Lambda^2}\rg-4\widehat
m_{\psi_{\pm}}^4\ln{m_{\widehat\psi_{\pm}}^2\over\Lambda^2}\rg
,\label{Vhic}\eea
where $\Lambda$ is a \emph{renormalization group} ({\sf\ftn RG})
mass scale. As we numerically verify the one-loop corrections have
no impact on our results. The absence of gauge interactions and of
a direct renormalizable coupling between $T$ and $\sni$ assists to
that direction -- cf.~\cref{talk,circ}. Based on $\Ve_{\rm MI}$,
we can proceed to the analysis of \FHI in the EF, employing the
standard slow-roll approximation \cite{review, lectures}. It can
be shown \cite{induced} that the results calculated this way are
the same as if we had calculated them using the non-minimally
coupled scalar field in the JF.

\subsection{The Inflationary Observables -- Requirements}\label{fhi2}

A successful inflationary scenario has to be compatible with a
number of observational requirements which are outlined in the
following.

\paragraph{3.2.1} The number of
e-foldings, $\widehat N_\star$, that the scale $k_\star=0.05/{\rm
Mpc}$ suffers during \FHI has to be adequate to resolve the
horizon and flatness problems of the standard Big Bag cosmology.
Assuming that \FHI is followed in turn by a decaying-particle,
radiation and matter domination and employing standard methods
\cite{nmi}, we can easily derive the required $\widehat{N}_\star$
for our model, with the result:
\begin{equation}  \label{Ntot}
\widehat{N}_\star\simeq19.4+2\ln{\what V_{\rm
MI}(\sg_\star)^{1/4}\over{1~{\rm GeV}}}-{4\over 3}\ln{\what V_{\rm
MI}(\sg_{\rm f})^{1/4}\over{1~{\rm GeV}}}+ {1\over3}\ln {T_{\rm
rh}\over{1~{\rm
GeV}}}+{1\over2}\ln{\fk(\sg_\star)\over\fk(\sg_{\rm f})^{1/3}},
\end{equation}
where $\sg_\star~[\se_\star]$ is the value of $\sg~[\se]$ when
$k_\star$ crosses the inflationary horizon. Also $\phi_{\rm
f}~[\se_{\rm f}]$ is the value of $\sg~[\se]$ at the end of \FHI
determined, in the slow-roll approximation, by the condition:
\beqs\beq \label{sr12} {\ftn\sf max}\{\what\epsilon(\sgf),\
|\what\eta(\sgf)|\}=1,\eeq where the slow-roll parameters read
\beq \label{sr1}\what \epsilon= {\mP^2\over2}\left(\frac{\Ve_{{\rm
MI},\se}}{\Ve_{\rm
MI}}\right)^2={\mP^2\over2J^2}\left(\frac{\Ve_{{\rm
MI},\phi}}{\Ve_{\rm MI}} \right)^2\simeq \frac{4}{3 \ft^2}\eeq and
\beq \label{sr2}\what\eta= m^2_{\rm P}~\frac{\Ve_{{\rm
MI},\se\se}}{\Ve_{\rm MI}}={\mP^2\over J^2}\left( \frac{\Ve_{{\rm
MI},\sg\sg}}{\Ve_{\rm MI}}-\frac{\Ve_{{\rm MI},\sg}}{\Ve_{\rm
MI}}{J_{,\sg}\over J}\right)\simeq\frac{4(1+\ft)}{3 \ft^2}\cdot
\eeq\eeqs
The termination of \FHI is triggered by the violation of the
$\epsilon$ criterion at a value of $\sg$ equal to $\sgf$, which is
calculated to be
\beqs\beq \what\epsilon\lf\sgf\rg=1\>\Rightarrow\> \sgf=\mP\lf(1 +
2/\sqrt{3})/\ck\rg^{1/2}, \label{sgap}\eeq
since the violation of the $\eta$ criterion occurs at
$\sg=\tilde\sg_{f}$ such that
\beq \what\eta\lf\tilde\sg_{\rm f}\rg=1\>\Rightarrow\>
\tilde\sgf=\mP\lf5 /3\ck\rg^{1/2}<\sgf. \label{sgap1}\eeq\eeqs

On the other hand, $\widehat N_\star$ can be calculated via the
relation
\begin{equation} \label{Nhi} \what{N}_\star=\:\frac{1}{m^2_{\rm
P}}\; \int_{\se_{\rm f}}^{\se_\star}\, d\se\:
\frac{\Vhi}{\Ve_{{\rm MI},\se}}= {1\over\mP^2}\int_{\phi_{\rm
f}}^{\phi{*}}\, d\sg\: J^2\frac{\Ve_{\rm MI}}{\Ve_{{\rm
MI},\phi}}\cdot
\end{equation}
Given that $\sgf\ll\sg_\star$, we can find a relation between
$\sg_\star$ and ${\Ne}_\star$ as follows
\beqs\beq \label{s*}
{\Ne}_\star\simeq{3\ck\over4\mP^2}\lf{\sg_\star^2-\sgf^2}\rg\>\Rightarrow\>
\sg_\star\simeq2\mP\sqrt{\Ne_\star/3\ck}.\eeq
Obviously, \FHI with \sub\ $\sg$'s can be achieved if
\beq \label{fsub}
\sg_\star\leq\mP~~\Rightarrow~~\ck\geq4\Ne_\star/3\simeq76
\eeq\eeqs
for $\Ne_\star\simeq52$. Therefore we need relatively large
$\ck$'s.

\paragraph{3.2.2} The amplitude $A_{\rm s}$ of the power spectrum of the curvature
perturbation generated by $\phi$ at the pivot scale $k_\star$ is
to be confronted with the data~\cite{wmap,plin}, i.e.
\begin{equation}  \label{Prob}
A^{1/2}_{s}=\: \frac{1}{2\sqrt{3}\, \pi\mP^3} \; \frac{\Ve_{\rm
MI}(\sex)^{3/2}}{|\Ve_{{\rm MI},\se}(\sex)|}=
{1\over2\pi\mP^2}\,\sqrt{\frac{\Vhi(\sg_\star)}{6\what\epsilon\,
(\sg_\star)}} \simeq4.685\cdot 10^{-5}. \eeq
Since the scalars listed in \Tref{tab3} are massive enough during
nSMI, the curvature perturbations generated by $\sg$ are solely
responsible for $A_{\rm s}$. Substituting \eqs{sr1}{s*} into the
relation above, we obtain
\beqs\beq \sqrt{\As}= \frac{\ld\mP^2 \ft(\sg_\star)^2}{8
\sqrt{2}\pi \ck^2 \sg_\star^2}
~~\Rightarrow~~\ld\simeq6\pi\sqrt{2\As}\ck/\Ne_\star.\eeq
Combining the last equality with \Eref{Prob}, we find that $\ld$
is to be proportional to $\ck$, for almost constant $\Ne_\star$.
Indeed, we obtain
\beq \ld\simeq{3.97\cdot10^{-4}\pi\ck/\Ne_\star}\>\Rightarrow\>
\ck\simeq41637\ld\>\>\>\mbox{for}\>\>\>\Ne_\star\simeq52.\label{lan}\eeq\eeqs

\paragraph{3.2.3} The (scalar) spectral index $n_{\rm s}$, its
running $a_{\rm s}$, and the scalar-to-tensor ratio $r$ must be
consistent with the fitting \cite{wmap,plin} of the observational
data, i.e.,
\begin{equation}  \label{obs}
\mbox{\ftn\sf (a)}\>\>\>\ns=0.96\pm0.014,\>\>\>\mbox{\ftn\sf
(b)}\>\>-0.0314\leq a_{\rm s}\leq0.0046
\>\>\>\mbox{and}\>\>\>\mbox{\ftn\sf (c)}\>\>r<0.11
\end{equation}
at 95$\%$ \emph{confidence level} (c.l.). The observable
quantities above can be estimated through the relations:
\beqs\baq \label{ns} && n_{\rm s}=\: 1-6\what\epsilon_\star\ +\
2\what\eta_\star\simeq1-{2/\what N_\star}-9/2\what N_\star^2, \>\>\> \\
&& \label{as} a_{\rm s}
=\:{2\over3}\left(4\what\eta_\star^2-(n_{\rm
s}-1)^2\right)-2\what\xi_\star\simeq-2\what\xi_\star\simeq{-2/ \what N^2_\star}+3/2\what N^3_\star,\>\>\> \\
 && \label{rt} r=16\what\epsilon_\star\simeq{12/\what N^2_\star},
\eaq\eeqs
where $\what\xi=\mP^4 {\Ve_{{\rm MI},\se} \Ve_{{\rm
MI},\se\se\se}/\Vhi^2}=
\mP\,\sqrt{2\what\epsilon}\what\eta_{,\phi}/ J+2\what
\eta\what\epsilon$. The variables with subscript $\star$ are
evaluated at $\phi=\phi_\star$ and \eqs{sr1}{sr2} have been
employed.

\subsection{The Effective Cut-off Scale}\label{fhi3}

As anticipated in \Eref{fsub}, the realization of nSMI with \sub\
$\phi$'s requires relatively large $\ck$'s. This fact may
\cite{cutoff} jeopardize the validity of the classical
approximation, on which the analysis of the inflationary behavior
is based. To see if this problem -- which is rather questionable
\cite{cutof, linde1} though -- insists here, we have to extract
the UV cut-off scale, $\Qef$, of the effective theory.

We first determine $\Qef$ analyzing the small-field behavior of
the model in EF along the lines of \cref{riotto}. The EF action
${\sf S}$ in \Eref{Saction1} along the path of \Eref{inftr} is
written as
\beqs \beq\label{S3} {\sf S}=\int d^4x \sqrt{-\what{
\mathfrak{g}}}\lf-\frac{1}{2}\mP^2 \rce +\frac12\,J^2
\dot\phi^2-\Ve_{\rm MI0}+\cdots\rg. \eeq
Given the form of $J$ in \Eref{cannor3b} an expansion of the
kinetic term in \Eref{S3} about zero is not doable. Therefore we
expand it about $\vev{\phi}=\mP/\sqrt{\ck}$ -- see \eqs{vevs}{ig}
-- and we find
\beq J^2
\dot\phi^2=6\ck\lf1-\frac{2\sqrt{\ck}\dph}{\mP}+\frac{3\ck\dph^2}{\mP^2}-
\frac{4\ck\sqrt{\ck}\dph^3}{\mP^3}+\frac{5\ck^2\dph^4}{\mP^4}-\cdots\rg\dot\phi^2,\label{exp1}\eeq
where $\dph=(\sg-M)$. Since there is no canonically normalized
leading kinetic term, we define the canonically normalized
inflaton at the SUSY vacuum $\dphi=\sqrt{6\ck}\dph$ -- see also
\Sref{lept} -- and we reexpress \Eref{exp1} in terms of $\dphi$,
with result
\beq\label{exp2} J^2
\dot\phi^2=\lf1-\sqrt{\frac{2}{3}}\frac{\dphi}{\mP}+\frac12\frac{\dphi^2}{\mP^2}-
\frac{\sqrt{2}}{3\sqrt{3}}\frac{\dphi^3}{\mP^3}+\frac{5}{36}\frac{\dphi^4}{\mP^4}-\cdots\rg\dot\dphi^2.\eeq
On the other hand, $\Vhio$ in \Eref{Vhio} can be expanded also in
terms of $\dphi$ as follows
\beq\label{Vexp}
\Vhio=\frac{\ld^2\mP^2}{6\ck^2}\dphi^2\lf1-\sqrt{\frac{3}{2}}\frac{\dphi}{\mP}+
\frac{25}{24}\frac{\dphi^2}{\mP^2}-\cdots\rg\cdot\eeq \eeqs
From the derived expressions in \eqs{exp2}{Vexp} we conclude that
$\Qef=\mP$ and therefore our model is valid up to $\mP$ as the
original Starobinsky model \cite{riotto}.

The resulting $\Qef$ represents essentially the
unitarity-violation scale \cite{cutoff} of the $\dph-\dph$
scattering process via $s$-channel graviton, $h^{\mu\nu}$,
exchange in the JF. The relevant vertex is $\ck\dph^2\Box  h/\mP$
-- with $h=h^\mu_{\mu}$ -- can be derived from the first term in
the r.h.s of \Eref{Sfinal} expanding the JF metric $g_{\mu\nu}$
about the flat spacetime metric $\eta_{\mu\nu}$ and the inflaton
$\phi$ about its v.e.v as follows:
\beq
g_{\mu\nu}\simeq\eta_{\mu\nu}+h_{\mu\nu}/\mP\>\>\>\mbox{and}\>\>\>\phi=\vev{\phi}
+\dph.\eeq
Retaining only the terms with two derivatives of the excitations,
the part of the lagrangian corresponding to the two first terms in
the r.h.s of \Eref{Sfinal} takes the form
\beqs\baq \nonumber \delta{\cal L}&=&-{\vev{\fk}\over4}{F}_{\rm
EH}\lf h^{\mu\nu}\rg+\lf\mP
\vev{f_{K,\phi}}+{\ck\dph\over2\mP}\rg\lf\Box h-
\partial_\mu \partial_\nu h^{\mu\nu}\rg\dph\ +\cdots\\ &=&-{1\over8}F_{\rm
EH}\lf \bar h^{\mu\nu}\rg+ \frac12\partial_\mu
\overline\dph\partial^\mu\overline\dph+\frac{1}{2\sqrt{2}}{\ck\over\mP}
\frac{\sqrt{\vev{\fk}}}{\vev{\bar{f}_K}}\,\overline\dph^2\,\Box
\bar h\, +\ \cdots,\label{L2}\eaq
where the function $F_{\rm EH}$, related to the the linearized
Einstein-Hilbert part of the lagrangian, reads
\beq {F}_{\rm EH}\lf h^{\mu\nu}\rg= h^{\mu\nu} \Box
h_{\mu\nu}-h\Box h+2\partial_\rho h^{\mu\rho}\partial^\nu
h_{\mu\nu}-2\partial_\nu h^{\mu\nu}\partial_\mu h\label{Leh}\eeq
and the JF canonically normalized fields $\bar h_{\mu\nu}$ and
$\overline\dph$ are defined by the relations
\beq
\overline\dph=\sqrt{\frac{\vev{\bar{f}_K}}{\vev{\fk}}}\dph\>\>\>\mbox{and}\>\>\>
{\bar h_{\mu\nu}\over\sqrt{2}}=
\sqrt{\vev{\fk}}\,h_{\mu\nu}+\frac{\mP\vev{f_{K,\phi}}}{\sqrt{\vev{\fk}}}\eta_{\mu\nu}\dph\>\>\>\mbox{with}\>\>\>
\bar{f}_K=3\mP^2 f_{K,\phi}^2.\label{Jcan}\eeq\eeqs
The interaction originating from the last term in the r.h.s of
\Eref{L2} gives rise to a scattering amplitude which is written in
terms of the center-of-mass energy $E$ as follows
\beq {\cal A}\sim\lf{E\over\Qef}\rg^2\>\>\>\mbox{with} \>\>\>\Qef=
{\mP\over3\sqrt{2}\ck}\frac{\vev{\bar{f}_K}}{\sqrt{\vev{\fk}}}=\mP,\eeq
where $\vev{\fk}=1/2$ and $\vev{\bar{f}_K}=3\ck$ and $\Qef$ is
identified as the UV cut-off scale in the JF, since ${\cal A}$
remains within the validity of the perturbation theory provided
that $E<\Qef$.

Although the expansions in \eqs{Vexp}{L2} are obtained for
$\phi\simeq\vev{\phi}$ and are not valid \cite{cutof} during nSMI,
we consider $\Qef$ as the overall UV cut-off scale of the model
since reheating is an unavoidable stage of the inflationary
dynamics \cite{riotto}. Therefore, the validity of the effective
theory implies \cite{cutoff}
\beq \label{Vl} \Vhi(\sg_\star)^{1/4}\ll\Qef
\>\>\>\mbox{with}\>\>\>\Qef=\mP,\eeq which is much less
restrictive than the corresponding condition applied in the models
of nMI with quartic scalar potential, where $\Qef$ turns out to be
equal to $\mP$ divided by the strength of the non-minimal coupling
to gravity -- cf. \cref{nmN,nmH,talk,riotto}.

\subsection{Numerical Results}\label{num1}

As can be easily seen from the relevant expressions above, the
inflationary dynamics of our model depends on the following
parameters:
$$\ld,\>\ck,\>\lm,\>k_{SS}=\kx,\>\ksh,\>\ksn,\>\mrh[i]\>\>\>\mbox{and}\>\>\>\Trh.$$
Recall that $M$ is related to $\ck$ via \Eref{ig}. Our results are
essentially independent of $\lm$ and $k$'s, provided that we
choose them so as $\what m^2_{h-}$ and $\what m_{s}^2$ in
\Tref{tab3} are positive for every allowed $\ld$. We therefore set
$\lm=10^{-6},\kx=\ksn=1$ and $\ksh=1.5$ throughout our
calculation. Moreover we take into account the contribution to
$\Vhi$, \Eref{Vhic}, only from the heaviest $\sni$ which is taken
to be $\mrh[3]=10^{14}~\GeV$ -- cf.~\Sref{num}. We also choose
$\Lambda\simeq10^{13}~\GeV$ so as the one-loop corrections in
\Eref{Vhic} vanish at the SUSY vacuum, \eqs{vevs}{ig}. Finally
$\Trh$ can be calculated self-consistently in our model as a
function of the inflaton mass, $\msn$ and the strength of the
various inflaton decays -- see \Sref{lept}. However, since the
inflationary predictions depend very weakly on $\Trh$ -- see
Eq.~(\ref{Ntot}) -- we prefer to take here a constant
$\Trh=6\cdot10^8~\GeV$ as suggested by our results on
post-inflationary evolution -- see \Sref{num}. Upon substitution
of $\Vhi$ from \Eref{Vhic} in \eqss{sr12}{Nhi}{Prob} we extract
the inflationary observables as functions of $\ck$, $\ld$ and
$\sg_\star$. The two latter parameters can be determined by
enforcing the fulfilment of \Eref{Ntot} and (\ref{Prob}), for
every chosen $\ck$. Our numerical findings are quite close to the
analytic ones listed in \Sref{fhi2} for the sake of presentation.

%%%%%%%%%%%%%%%%%%%%%%%%%%%%%%%%%%%%%%%%%%%%%%%%%%%%%%%%%%%%%%%%%%%%
\begin{figure}[!t]\vspace*{-.26in}
\begin{center}
\epsfig{file=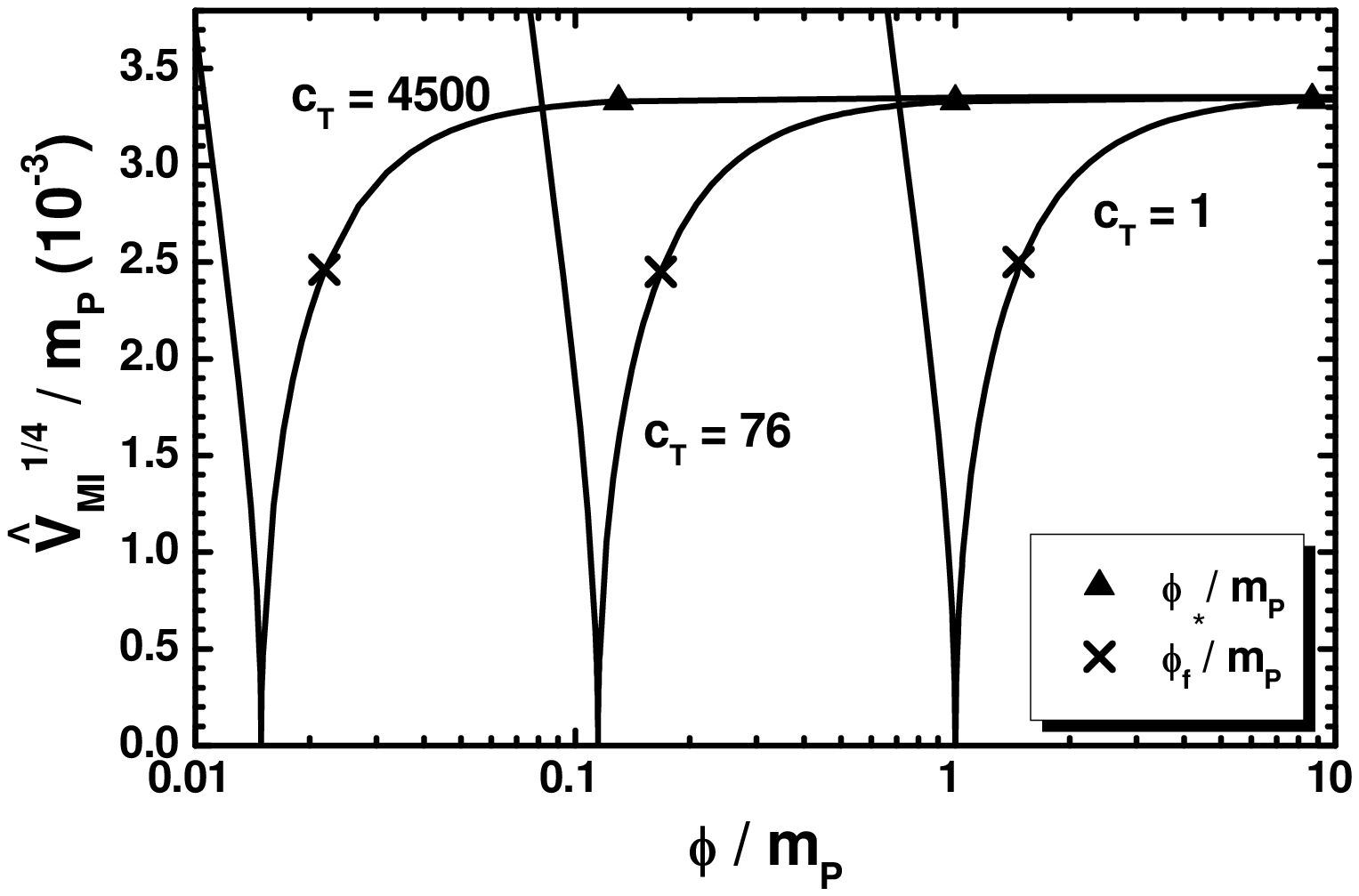,height=3.65in,angle=-90}\ec
\vspace*{-.15in}\hfill \vchcaption[]{\sl \small The inflationary
scale $\Vhi^{1/4}$ as a function of $\sg$ for
$\ld=2.26\cdot10^{-5}$ and $M=\mP$ ($\ck=1$) or
$\ld=1.7\cdot10^{-3}$ and $M/\mP=0.115$ ($\ck=76$) or $\ld=0.1$
and $M/\mP=0.015$ ($\ck=4500$). The values corresponding to $\sgx$
and $\sgf$ are also depicted.}\label{fig3}
\end{figure}
%%%%%%%%%%%%%%%%%%%%%%%%%%%%%%%%%%%%%%%%%

The importance of the two extra variables ($M$ and $\ck$) -- in
\eqss{Whi}{Kol}{fdef} -- compared to the Cecotti model
\cite{linde, eno7,kehagias,zavalos} in reducing $\sgx$ below $\mP$
can be easily inferred from \Fref{fig3}. We there depict
$\Vhi^{1/4}$ as a function of $\sg$ (both normalized to $\mP$) for
$\ld=2.26\cdot10^{-5}$ and $\ck=1$ or $\ld=0.0017$ and $\ck=76$ or
$\ld=0.1$ and $\ck=4500$ -- the last value saturates an upper
bound on $\ck$ derived in \Sref{secmu}. Note that for $\ck=1$ (or
$\mpq=1$) our result matches that of the original Starobinsky
model \cite{defelice,eno5} -- with the mass scale appearing in
that model being replaced by $\ld\mP\simeq2.2\cdot10^{13}~\GeV$.
Increasing $\ck$, $\ld$ increases too, whereas $\sgx$ and $M$
decrease and for $\ck\geq76$, $\sgx$ becomes \sub. On the other
hand, we have to clarify that the corresponding values of the
inflaton in the EF remain \trns, since integrating the first
equation in \Eref{cannor3b} and using \eqs{s*}{sgap} we find:
\beq \se=\se_{\rm
c}+\sqrt{6}\mP\ln\lf\sg/M\rg~~\Rightarrow~~\left\{\bem
%\begin{array}{rl}
%
\se_\star-\se_{\rm c}\simeq\sqrt{6}\mP\ln 2(\Ne_\star/3)^{1/2}
\hfill \cr
\se_{\rm f}-\se_{\rm c}\simeq\sqrt{6}\mP
\ln(1+2/\sqrt{3})^{1/2}.\hfill \cr \eem
%\end{array}
\right. \label{se1}\eeq
where $\se_{\rm c}$ is a constant of integration. E.g., setting
$\se_{\rm c}=0$, we obtain $\se_\star=5.3\mP$ and $\se_{\rm
f}=0.94\mP$ for any $\ck$ -- with constant $\Ne_\star$. We do not
consider this result as an upset of our proposal, since the
inflaton field defined in the JF enters $\Whi$ and $K$. Therefore,
possible corrections from non-renormalizable terms, which may be
avoided for \sub\ values of inflaton, are applied in this frame,
which is mostly considered as the physical frame.

%%%%%%%%%%%%%%%%%%%%%%%%%%%%%%%%%%%%%%%%%%%%%%%%%%%%%%%%%%%%%%%%%%%%%
\begin{figure}[!t]\vspace*{-.12in}
\hspace*{-.19in}
\begin{minipage}{8in}
\epsfig{file=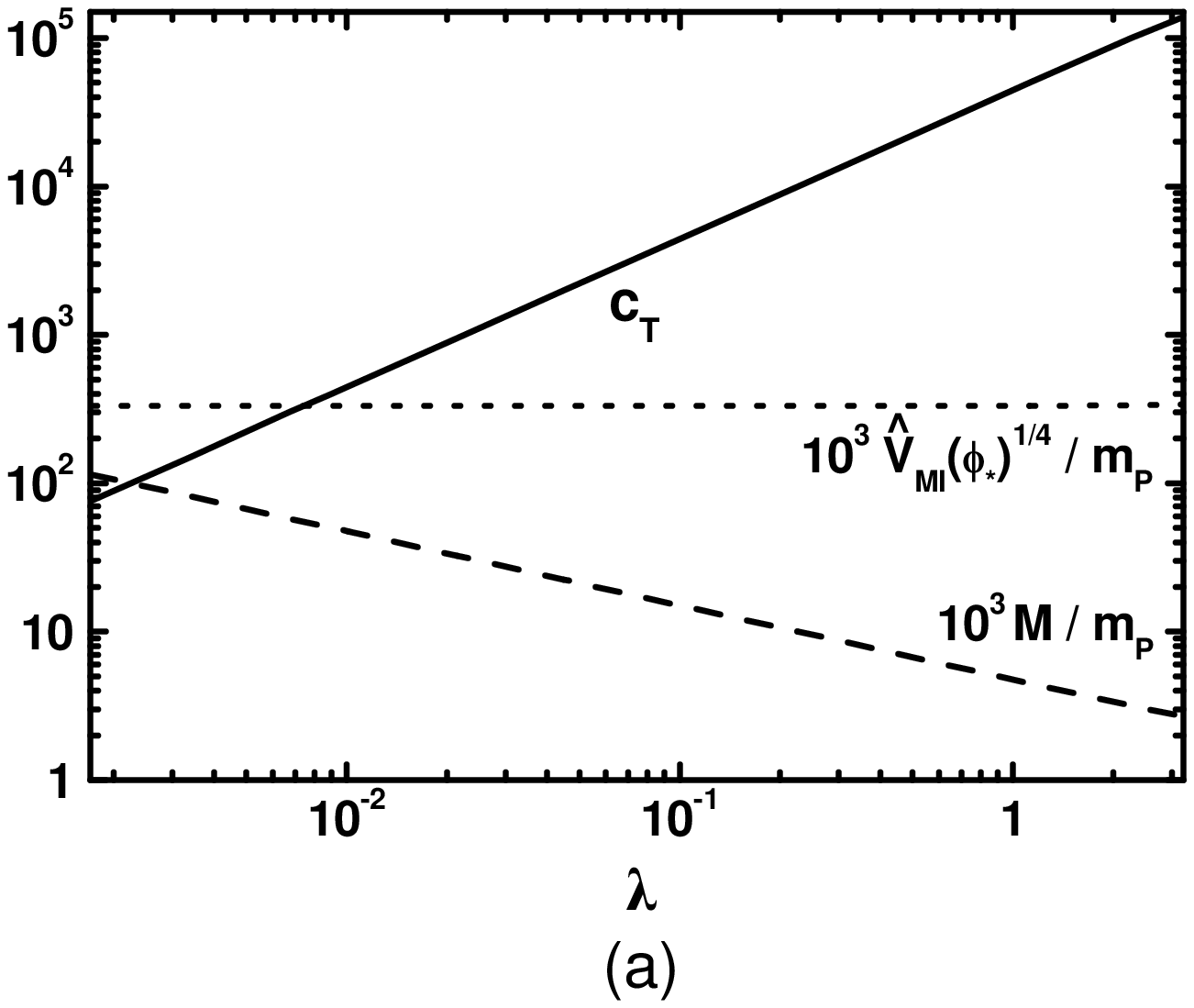,height=3.6in,angle=-90}
\hspace*{-1.2cm}
\epsfig{file=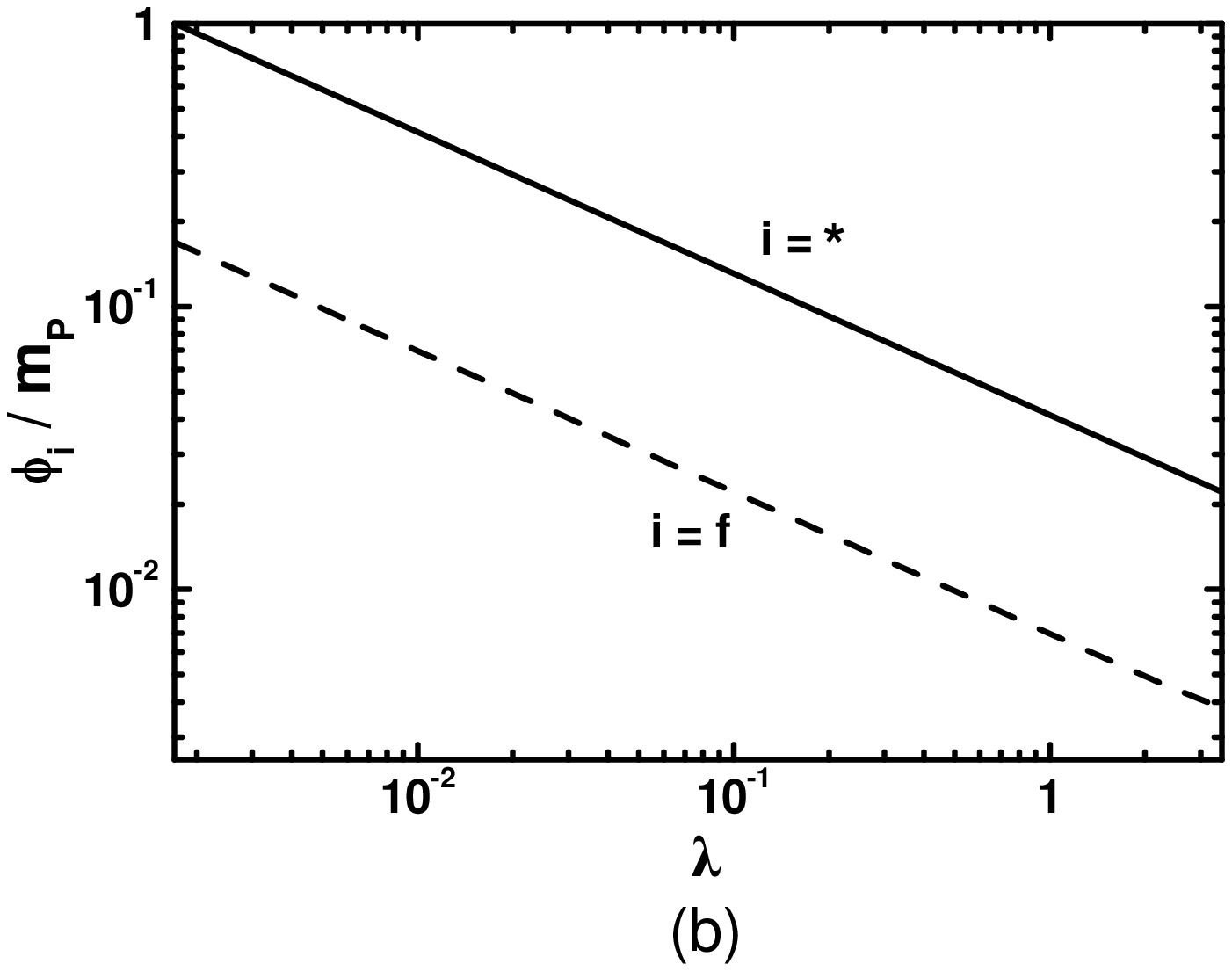,height=3.6in,angle=-90} \hfill
\end{minipage}
\hfill \vchcaption[]{\sl\small  The allowed by Eqs.~(\ref{Ntot}),
(\ref{Prob}) and (\ref{Vl}) values of $\ck$ (solid line),
$10^3\mpq$ (dashed line) and $10^3\Vhi(\sg_\star)^{1/4}/\Qef$
(dotted line) [$\sg_{\rm f}$ (solid line) and $\sg_\star$ (dashed
line)] versus $\ld$ (a) [(b)] for $\kx=1,~\lm=10^{-6}$,
$\mrh[3]=10^{14}~\GeV$ and $\Trh=6\cdot10^{8}~\GeV$. }\label{fig1}
\end{figure}

%%%%%%%%%%%%%%%%%%%%%%%%%%%%%%%%%%%%%%%%%%

From \Fref{fig3} we also infer that $\Vhi^{1/4}/\mP$ remains
almost constant during nSMI. Indeed, if we plug \eqs{s*}{lan} into
\Eref{Vhio}, we obtain
\beq \label{V*} \Vhio(\sgx)^{1/4}/\mP\simeq\lf
3\pi\sqrt{2\As}/\Ne_\star\rg^{1/2}\simeq0.0033\ll1.\eeq
This result is more explicitly displayed in \Fref{fig1} too, where
we draw the allowed values of $\ck$ (solid line), $10^3\mpq$
(dashed line) and $10^3\Vhi(\sg_\star)^{1/4}/\mP$ (dotted line)
[$\sg_{\rm f}$ (solid line) and $\sg_\star$ (dashed line)] versus
$\ld$ (a) [(b)]. The lower bound of the depicted lines comes from
the saturation of \Eref{fsub} whereas the upper bound originates
from the perturbative bound on $\ld$,
$\ld\leq\sqrt{4\pi}\simeq3.54$. In \sFref{fig1}{a} we see that
\Eref{Vl} is readily satisfied along the various curves and we can
verify our analytic estimation in \Eref{lan}. Moreover, the
variation of $\sg_{\rm f}$ and $\sg_\star$ as a function of $\ld$
-- drawn in \sFref{fig1}{b} -- is consistent with \eqs{sgap}{s*}.
The overall allowed parameter space of our model is
\beqs\beq\label{res1} 76\lesssim
\ck\lesssim1.5\cdot10^5,\>\>\>0.11\gtrsim
\mpq\gtrsim0.002\>\>\>\mbox{and}\>\>\>1.7\cdot10^{-3}\lesssim
\ld\lesssim3.54\>\>\>\mbox{for}\>\>\> \Ne_\star\simeq52.\eeq
Letting $\ld$ or $\ck$ vary within its allowed region in
\Eref{res1}, we obtain
\beq\label{res} 0.961\lesssim \ns\lesssim0.963,\>\>\>-7.4\lesssim
{\as/10^{-4}}\lesssim-6.7\>\>\>\mbox{and}\>\>\>4.2\gtrsim
{r/10^{-3}}\gtrsim3.8,\eeq
whereas the masses of the various scalars in \Tref{tab3} remain
well above $\Hhi$ both during and after nSMI for the selected
$\kx,\lm$ and $\mrh[3]$. E.g., for $\sg=\sgx$ and $\ck=150$, we
obtain
\beq\label{res3} \lf \what m_{\th}^2, \what m_{s}^2, \what
m_{h-}^2, \what m_{h+}^2, \what m_{
3\nu^c}^2\rg/\Hhi^2\simeq(4,905,342,342,282). \eeq \eeqs
Clearly, the predicted $\as$ and $r$ lie within the allowed ranges
given in \sEref{obs}{b} and \sEref{obs}{c} respectively, whereas
$\ns$ turns out to be impressively close to its central
observationally favored value -- see \sEref{obs}{a}. Therefore,
the inclusion of extra parameters, compared to the Cecotti model
\cite{linde,eno7,kehagias,zavalos}, does not affect the successful
predictions on the inflationary observables.

\section{The $R$ Symmetry and the $\mu$ Problem of MSSM}
\label{secmu}

A byproduct of the $R$ symmetry associated with our model is that
it assists us to understand the origin of $\mu$ term of MSSM. To
see how this works, in \Sref{secmu2}, we estimate the $\mu$
parameter and, in \Sref{pheno}, we control its compatibility with
phenomenologically acceptable values obtained in the context of
the \emph{Constrained MSSM} ({\ftn\sf CMSSM}) \cite{mssm}.

\subsection{Generation of the $\mu$ Term of MSSM}\label{secmu2}

First we write the part of the scalar potential which includes the
SSB terms corresponding to $\Whi$ in \Eref{Whi}. We have
\beq V_{\rm soft}= \lf\ld A_\ld S T^2 +\lm A_\mu S \hu\hd +
B_{iN^c}\mrh[i]\ssni\ssni- {\rm a}_{S}S\ld M^2 + {\rm h. c.}\rg+
m_{\al}^2\left|\Phi^\al\right|^2, \label{Vsoft} \eeq
where $m_{\al}, A_\ld, A_\mu, B_{iN^c}$ and $\aS$ are SSB mass
parameters. Rotating $S$ in the real axis by an appropriate
$R$-transformation, choosing conveniently the phases of $\Ald$ and
$\aS$ so as the total low energy potential $V_{\rm tot}=V_{\rm
SUSY}+V_{\rm soft}$ to be minimized -- see \Eref{VF} -- and
substituting in $V_{\rm soft}$ the SUSY v.e.vs of $T,\hu, \hd$ and
$\ssni$ from \Eref{vevs} we get
\beq \vev{V_{\rm tot}(S)}= \frac{\ld^2\,M^2}{3\ck\mP^2}S^2\lf\ck
M^2-\frac13S^2\rg-2\ld\am\mgr M^2 S, \label{Vol} \eeq
where we set $|A_\ld| + |{\rm a}_{S}|=2\am\mgr$ with $\mgr$ being
the $\Gr$ mass and $\am>0$ a parameter of order unity which
parameterizes our ignorance for the dependence of $|A_\ld|$ and
$|{\rm a}_{S}|$ on $\mgr$. Making use of the induced-gravity
condition in \Eref{con}, we can write the extremum condition for
$\vev{V_{\rm tot}(S)}$ as follows
\beqs\beq \label{Vs} \frac{d}{d S} \vev{V_{\rm
tot}(S)}=0~~\Rightarrow~~-\frac{2\ld^2\mP^2}{3\ck^2}\lf
\frac23\frac{S^3}{\mP^2}-S+\frac3\ld\ck\am\mgr\rg=0\,.\eeq
For $\vev{S}\ll\mP$, one of the solutions of the last equation is
\beq \label{vevS} \vev{S}\simeq 3\ck\am\mgr
/\ld\,\simeq1.25\cdot10^5\am\mgr,\eeq\eeqs
where we employ \Eref{lan} which yields $\ld$ as a function of
$\ck$. At this $S$ value, $\vev{V_{\rm tot}(S)}$ develops a
minimum since
\beqs\beq \label{Vss}\frac{d^2}{d S^2} \vev{V_{\rm
tot}(S)}=\frac{2 \ld^2}{3 \ck^2} \lf \mP^2 - 2 S^2\rg\eeq
becomes positive for $\mgr\ll\mP$. Indeed, inserting \Eref{vevS}
into \Eref{Vss} we obtain the constraint
\beq
3\sqrt{2}\ck\frac{\am\mgr}{\ld\mP}\leq1~~\Rightarrow~~\mgr\leq\frac{\ld\mP}{3\sqrt{2}\am\ck}
\simeq1.38\cdot10^{13}~\GeV,\eeq\eeqs
which is comfortably fulfilled in the case of the low scale SUSY.
The other two solutions of \Eref{Vs} violate this bound. The
generated $\mu$ parameter from the second term in the r.h.s of
\Eref{Whi} is
\beq\mu =\lm \vev{S} \simeq3\lm\ck\am\mgr/\ld \simeq
1.25\cdot10^5\lm\am\mgr\,.\label{mu}\eeq
Note that $\lm$ (and so $\mu$) may have either sign without any
essential alteration in the stability analysis of the inflationary
system -- see \Tref{tab3}. Thanks to the magnitude of the
proportionality constant, any ratio $|\mu|/\am\mgr\lesssim2.5$ is
accessible for the $\lm$ values allowed by \Eref{lm} with $\ksh$
of order unity. Ergo, the resulting $\mu$ is $\ld$ independent, in
sharp contrast to the originally proposed scheme in \cref{dvali},
and no hierarchy of the type $\mu\ll\mgr$ is required -- see e.g.
the second row of \Tref{tab} below where $\mu/\am\mgr\simeq2.13$.

Obviously the proposed resolution of the $\mu$ problem of MSSM
relies on the existence of non-zero $\Ald$ and/or $\aS$. These
issues depend on the adopted model of SSB. We single out the
following cases:

\begin{itemize}

\item[{\sf \ftn (i)}] If we wish to be fully consistent the
no-scale structure of $K$ and suppose that the modulus, $z$, which
is responsible for the SSB, is contained (somehow) in the
logarithm of \Eref{Kol}, $K$ is of the ``sequestered-sector'' form
\cite{randal} and has the property that it generates no tree-level
SSB scalar masses for the visible-sector fields and vanishing
trilinear coupling constants. In this case the anomaly-mediated
SSB \cite{randal, anshafi} is the dominant mechanism for obtaining
$\Ald\neq0$ and/or $\aS\neq0$. Since the involved superfields $T$
and $S$ are $\Gsm$ singlets, we expect $\Ald=0$. However,
according to the superconformal formalism, $M^2$ can be rescaled
as $M^2\varphi^2$ (where $\varphi$ is a superconformal
compensator) and, in the presence of SSB, a non vanishing
$\aS=2\mgr$ comes out.

\item[{\sf \ftn (ii)}] If we decide to deviate from the no-scale
form of $K$ in \Eref{Kol}, we can suppose that $z$ is not
contained in the logarithm, and has an almost canonical \Ka\
\cite{nilles, buch}. In a such circumstance, both $\Ald$ and $\aS$
are expected to be non-zero, as in the gravity-mediated SSB
\cite{nilles}, giving rise again to $\vev{S}\neq0$.

\end{itemize}

In both cases above, our \sup\ in \Eref{Wol} has to be extended by
a SSB sector which should ensure the successful stabilization of
$z$ -- cf.~\cref{buch,randal,olivegr}. We expect that these terms
do not disturb the inflationary dynamics. Alternatively, the $\mu$
problem can be resolved \cite{rsym} by imposing a Peccei-Quinn
symmetry which is broken spontaneously at an intermediate scale by
the v.e.vs of two $\Gsm$ singlets which enter the supepotential
via non-renormalizable terms. This scheme, already adopted, e.g.,
in \cref{nmH,rob}, can be applied as first realized in \cref{rsym}
in the case (ii) above and somehow modified in the case (i).

Let us clarify, finally, that the due hierarchy in \Eref{lm}
between $\lm$ and $\ld$, is the inverse to that imposed in the
models \cite{dvali} of FHI, where $S$ plays the role of inflaton
and $T,~\hu$ and $\hd$ are confined at zero -- playing the role of
the waterfall fields. This is because, at the end of FHI, the mass
squared of $T$ becomes negative for $S<M/\sqrt{2}$ and the mass
matrix squared of the scalars $\hu-\hd$ develop a negative
eigenvalue for $S<M\sqrt{\ld /2\lm}$. Consequently, the correct
cosmological scenario can be attained if we ensure that, at the
end of FHI, $T$ acquires its v.e.v, while $\hu$ and $\hd$ remain
equal to zero. To this end we demand \cite{dvali} $\lm>\ld$ so as
the tachyonic instability in the $T$ direction occurs first, and
$T$ start evolving towards its v.e.v, whereas $\hu$ and $\hd$
continue to be confined to zero. In our case, though, $|T|$ is the
inflaton while $S$ and the $\hu-\hd$ system are safely stabilized
at the origin both during and after the end of nSMI. Therefore,
$|T|$ is led at its vacuum whereas $\hu$ and $\hd$ take their
non-vanishing v.e.vs during the electroweak phase transition
triggered by radiative corrections.

\subsection{Connection With the MSSM Phenomenology}\label{pheno}

Taking advantage from the updated investigation of the parameter
space of CMSSM in \cref{mssm} we can easily verify that the $\mu$
and $\mgr$ values satisfying \Eref{mu} are consistent with the
values required by the analyses of the low energy observables of
MSSM. We concentrate on CMSSM which is the most predictive,
restrictive and well-motivated version of MSSM, employing the free
parameters:
%\begin{equation}
$$\sign\mu,~~\tan\beta=\vev{\hu}/\vev{\hd},~~\mg,~~m_0,~~\mbox{and}~~A_0.$$
%\nonumber
%\end{equation}
Here $\sign\mu$ is the sign of $\mu$ and the three last mass
parameters denote the common gaugino mass, scalar mass, and
trilinear coupling constant, respectively, defined at a high scale
which is determined by the unification of the gauge coupling
constants. The parameter $|\mu|$ is not free, since it is computed
at low scale enforcing the conditions for the electroweak symmetry
breaking. The values of these parameters can be tightly restricted
imposing a number of cosmo-phenomenological constraints. Some
updated results are recently presented in \cref{mssm}, where we
can also find the best-fit values of $|A_0|$, $m_0$ and $|\mu|$
listed in \Tref{tab}.  We see that there are four allowed regions
characterized by the specific mechanism for suppressing the relic
density of the lightest sparticle which can act as dark matter. If
we identify $m_0$ with $\mgr$ and $|A_0|$ with $|A_\ld|=|\aS|$ we
can derive first $\am$ and then the $\lm$ values which yield the
phenomenologically desired $|\mu|$ -- see the two rightmost
columns in \Tref{tab}. Here we assume that renormalization effects
are negligible. Since the required $\lm$'s are compatible with
\Eref{lm} for $\ksh=1.5$, we conclude that the whole inflationary
scenario can be successfully combined with CMSSM. On the other
hand, only the regions which become consistent with the dark
matter requirements thanks to $A/H$ funnel and $\tilde \chi^\pm_1$
coannihilation can be consistent with the gravitino limit on
$\Trh$ -- see \Sref{cont1}. Indeed, in these cases
$\mgr\simeq9~\TeV$ and so, the unstable $\Gr$ becomes
cosmologically safe with the presented in \Tref{tab2} $\Trh$
values, necessitated for satisfactory leptogenesis.

\renewcommand{\arraystretch}{1.25}
\begin{table} \bec
\begin{tabular}{|c|c|c|c||c|c|}\hline
{\sc CMSSM Region}&$|A_0| (\TeV)$&$m_0 (\TeV)$&$|\mu|
(\TeV)$&$\am$&$\lm (10^{-6})$\\
\hline\hline
$A/H$ Funnel &$9.9244$&$9.136$&$1.409$&$1.086$&$1.071$\\
$\tilde\tau$ Coannihilation &$1.2271$&$1.476$&$2.621$&$0.831$&$16.11$\\
$\tilde t$ Coannihilation  &$9.965$&$4.269$&$4.073$&$2.33$&$3.081$\\
$\tilde \chi^\pm_1$ Coannihilation  &$9.2061$&$9.000$&$0.983$&$1.023$&$0.805$\\
\hline
\end{tabular}
\end{center}
\caption[]{\sl\small The required $\lm$ values for $m_0=\mgr$,
$|A_\ld|=|\aS|=|A_0|$ and $\ksh=1.5$ which render our model
compatible with the best-fit points of the CMSSM as found in
\cref{mssm}.} \label{tab}
\end{table}\renewcommand{\arraystretch}{1.}

\section{Non-Thermal Leptogenesis and Neutrino Masses}\label{pfhi}

We below specify how our inflationary scenario makes a transition
to the radiation dominated era (\Sref{lept}) and give an
explanation of the observed BAU (\Sref{lept1}) consistently with
the $\Gr$ constraint and the low energy neutrino data
(\Sref{lept2}). Our results are summarized in \Sref{num}.

\subsection{The Inflaton Decay}\label{lept}

When \FHI is over, the inflaton continues to roll down towards the
SUSY vacuum, \Eref{vevs}. Soon after, it settles into a phase of
damped oscillations around the minimum of $\Vhio$ -- note that
$\th$ is stabilized during and after \FHI at the origin and so, it
does not participate neither into inflationary nor to
post-inflationary dynamics. The (canonically normalized) inflaton,
$\dphi=\sqrt{6\ck}\dph$ -- see, also, \Sref{fhi3} --, acquires
mass which is given by
\beq \label{masses} \msn=\left\langle\Ve_{\rm
MI0,\se\se}\right\rangle^{1/2}= \left\langle \Ve_{\rm
MI0,\sg\sg}/J^2\right\rangle^{1/2}={\ld\mP/\sqrt{3}\ck}\simeq3\cdot10^{13}~\GeV,\eeq
where we make use of \Eref{lan} in the last step. Since
\Eref{VJe3} implies $\vev{K_{\Al\Aal}}=1$ for
$\vev{\xsg}=1/\sqrt{\ck}$ -- see \eqss{dimlss}{vevs}{ig} --, the
EF canonically normalized fields $\Phi^A$ in \Eref{fas} are not
distinguished from the JF ones at the SUSY vacuum.

The decay of $\dphi$ is processed through the following decay
channels:

\paragraph{5.1.1 Decay channel into {\small $\sni$}'s. } The lagrangian which
describes these decay channels arises from the part of the SUGRA
langrangian \cite{nilles} containing two fermions. In particular,
\beqs\bea \nonumber {\cal L}_{\dphi\to
\sni}&=&-\frac12e^{K/2\mP^2}W_{,N_i^cN^c_i}\sni\sni\ +{\rm
h.c.}\,=
\frac{3}{2}\frac{M}{\mP}\ck^{1/2}\dph\ \sni\sni+\cdots\\
&=&\ld_{iN^c} \dphi\
\sni\sni+\cdots\>\>\>\mbox{with}\>\>\>\ld_{iN^c}={\sqrt{3}M_{iN^c}}/{2\sqrt{2}\mP},\label{Lnu}\eea
where an expansion around $\vev{\phi}$ is performed in order to
extract the result above. We observe that although there is not
direct coupling between $T$ and $N_i^c$ in $\Whi$ -- recall that
we assume that the third term in the r.h.s of \Eref{Whi}  prevails
over the last one --, an adequately efficient decay channel
arises, which gives rise to the following decay width
\beq
\GNsn=\frac{1}{16\pi}\ld_{iN^c}^2\msn\lf{1-4M_{iN^c}^2/\msn^2}\rg^{3/2},
\label{Gpq}\eeq\eeqs
where we take into account that $\dphi$ decays to identical
particles.

\paragraph{5.1.2. Decay channel into {\small $\hu$} and {\small $\hd$}.}  The lagrangian term which describes
the relevant interaction comes from the F-term SUGRA scalar
potential in \Eref{Vsugra}. Namely, we obtain
\beqs\bea\nonumber {\cal L}_{\dphi\to
\hu\hd}&=&-\frac12e^{K/\mP^2}K^{SS^*}\left|W_{S}\right|^2=-\frac12\ld\lm\lf
\phi^2-M^2\rg H_u^*H_d^*\ +\cdots\\ &=&-\ld_{H} \msn\dphi
H_u^*H_d^*\
+\cdots\>\>\>\mbox{with}\>\>\>\ld_{H}={\lm}/{\sqrt{2}}.\label{Lh}\eea
This interaction gives rise to the following decay width
\beq \Ghsn=\frac{2}{8\pi}\ld_{H}^2\msn, \label{Ghh}\eeq\eeqs
where we take into account that $\hu$ and $\hd$ are $SU(2)_{\rm
L}$ doublets. \Eref{lm} facilitates the reduction of $\Ghsn$ to a
level which allows for the decay mode into $\sni$'s playing its
important role for nTL.

\paragraph{5.1.3. Three-particle decay channels.} Focusing on the same part of the SUGRA
langrangian \cite{nilles} as in the paragraph 5.1.1, for a typical
trilinear superpotential term of the form $W_y=yXYZ$ -- cf.
\Eref{wmssm} --, where $y$ is a Yukawa coupling constant, we
obtain the interactions described by
\beqs\bea  \nonumber {\cal L}_{\dphi\to XYZ} &=&
-\frac12e^{K/2\mP^2}\lf
W_{y,YZ}\psi_{Y}\psi_{Z}+W_{y,XZ}\psi_{X}\psi_{Z}+
W_{y,XY}\psi_{X}\psi_{Y}\rg+{\rm h.c.}\, \\ &=&
\ld_y{\dphi\over\mP}\lf X\psi_{Y}\psi_{Z}+Y\psi_{X}\psi_{Z}+
Z\psi_{X}\psi_{Y}\rg+{\rm
h.c.}\,\>\>\>\mbox{with}\>\>\>\ld_y=\sqrt{3/2}({y}/{2}),\>\>\>
\label{Lxyz}\eea
where $\psi_X, \psi_{Y}$ and $\psi_{Z}$ are the chiral fermions
associated with the superfields $X, Y$ and $Z$ whose the scalar
components are denoted with the superfield symbol. Working in the
large $\tb$ regime which yields similar $y$'s for the 3rd
generation, we conclude that the interaction above gives rise to
the following 3-body decay width
\beq \Gysn={14 n_{\rm f}\over512\pi^3}\ld_y^2{\msn^3\over\mP^2},
\label{Gpq1}\eeq\eeqs
where for the third generation we take $y\simeq(0.4-0.6)$,
computed at the $\msn$ scale, and $n_{\rm f}=14$ [$n_{\rm f}=16$]
for $\msn<\mrh[3]$ [$\msn>\mrh[3]$] -- summation is taken over
$SU(3)_{\rm c}$ and $SU(2)_{\rm L}$ indices.

Since the decay width of the produced $\sni$ is much larger than
$\Gsn$ the reheating temperature, $\Trh$, is exclusively
determined by the inflaton decay and is given by \cite{quin}
\beq \label{Trh} \Trh=
\left(72\over5\pi^2g_*\right)^{1/4}\sqrt{\Gsn\mP}
\>\>\>\mbox{with}\>\>\>\Gsn=\GNsn+\Ghsn+\Gysn,\eeq
where $g_*\simeq228.75$ counts the effective number of
relativistic degrees of freedom of the MSSM spectrum at the
temperature $T\simeq\Trh$. Let us clarify here that in our models
there is no decay of a scalaron as in the original (non-SUSY)
\cite{R2,smR2} Starobinsky inflation and some \cite{ketov} of its
SUGRA realizations; thus, $\Trh$ in our case is slightly lower
than that obtained there. Indeed, spontaneous decay of the
inflaton to scalars takes place only via three-body interactions
which are suppressed compared to the two-body decays of scalaron.
On the other hand, we here get also $\Ghsn$ in \Eref{Ghh}, due to
explicit coupling of $\dphi$ into $\hu$ and $\hd$, which can be
kept at the same level with $\Gysn$ due to the rather low $\lm$'s
required here -- see \Eref{lm}.

\subsection{Lepton-Number and Gravitino Abundances}\label{lept1}

The mechanism of nTL \cite{lept} can be activated by the
out-of-equilibrium decay of the $N^c_{i}$'s produced by the
$\dphi$ decay, via the interactions in \Eref{Lnu}. If
$\Trh\ll\mrh[i]$, the out-of-equilibrium condition \cite{baryo} is
automatically satisfied. Namely, $N^c_{i}$ decay into (fermionic
and bosonic components of) $H_u$ and $L_i$ via the tree-level
couplings derived from the last term in the r.h.s of
Eq.~(\ref{wmssm}). The resulting -- see \Sref{lept2} --
lepton-number asymmetry $\ve_i$ (per $N^c_i$ decay) after
reheating can be partially converted via sphaleron effects into
baryon-number asymmetry. In particular, the $B$ yield can be
computed as
\beq {\ftn\sf
(a)}\>\>\>Y_B=-0.35Y_L\>\>\>\mbox{with}\>\>\>{\ftn\sf
(b)}\>\>\>Y_L=2{5\over4}
{\Trh\over\msn}\sum_{i=1}^3{\GNsn\over\Gsn}\ve_i\cdot\label{Yb}\eeq
The numerical factor in the r.h.s of \sEref{Yb}{a} comes from the
sphaleron effects, whereas the one ($5/4$) in the r.h.s of
\sEref{Yb}{b} is due to the slightly different calculation
\cite{quin} of $\Trh$ -- cf.~\cref{baryo}.

The required for successful nTL $\Trh$ must be compatible with
constraints on the $\Gr$ abundance, $Y_{\Gr}$, at the onset of
\emph{nucleosynthesis} (BBN). This is estimated to be
\cite{brand,kohri}:
\beq\label{Ygr} Y_{\Gr}\simeq 1.9\cdot10^{-22}\Trh/\GeV, \eeq
where we assume that $\Gr$ is much heavier than the gauginos of
MSSM. Let us note that non-thermal $\Gr$ production within SUGRA
is \cite{Idecay} also possible but strongly dependent on the
mechanism of SSB. It can be easily suppressed
\cite{eluding,spontaneous} when a tiny mixing arises between the
inflaton and the field responsible for SSB provided that the mass
of the latter is much lower than the inflationary scale.
Therefore, we here prefer to adopt the conservative estimation of
$Y_{\Gr}$ in \Eref{Ygr}.

Both \eqs{Yb}{Ygr} calculate the correct values of the $B$ and
$\Gr$ abundances provided that no entropy production occurs for
$T<\Trh$. This fact can be achieved if the Polonyi-like field $z$
decays early enough without provoking a late episode of secondary
reheating. In both cases of \Sref{secmu}, $z$ is expected to be
displaced from its true minimum to lower values due to large mass
that it acquires during nSMI. In the course of the
decaying-inflaton period which follows nSMI, $z$ adiabatically
tracks an instantaneous minimum \cite{adiabatic} until the Hubble
parameter becomes of the order of its mass. Successively it starts
to oscillate about the true SUSY breaking minimum and may or may
not dominate the Universe, depending on the initial amplitude of
the coherent oscillations. The domination may be eluded in a very
promising scenario \cite{olivegr,buch} which can be constructed
assuming that $z$ is strongly stabilized through a large enough
coupling in a higher order term of \Ka, similar to that used for
the stabilization of $S$ -- see \Eref{Kol}. A subsequent
difficulty is the possible over-abundance of the LSPs which are
produced by the $z$ decay. From that perspective, it seems that
the case (ii) -- cf.~\cref{olivegr, adiabatic} -- is more
tolerable than the case (i) -- see \cref{nsdecay}.

\subsection{Lepton-Number Asymmetry and Neutrino
Masses}\label{lept2}

As mentioned in \Sref{lept1}, the decay of $\sni$ emerging from
the $\dphi$ decay, can generate \cite{resonant1}  a lepton
asymmetry $\ve_i$ caused by the interference between the tree and
one-loop decay diagrams, provided that a CP-violation occurs in
$h_{ijN}$'s -- see \Eref{wmssm}. The produced $\ve_i$ can be
expressed in terms of the Dirac mass matrix of $\nu_i$, $m_{\rm
D}$, defined in the $\sni$-basis, as follows:
\beqs\beq\ve_i =\sum_{j\neq i}{
\im\left[(\mD[]^\dag\mD[])_{ij}^2\right]\over8\pi\vev{H_u}^2(\mD[]^\dag\mD[])_{ii}}
\bigg( F_{\rm S}\lf x_{ij}\rg+F_{\rm V}(x_{ij})\bigg) ,
\label{el}\eeq where $x_{ij}:={\mrh[j]/\mrh[i]}$,
$\vev{H_u}\simeq174~\GeV$, for large $\tan\beta$ and the functions
$F_{\rm V, S}$ read \cite{resonant1} \beq F_{\rm V}\lf
x\rg=-x\ln\lf1+ x^{-2}\rg\>\>\>\mbox{and}\>\>\>F_{\rm S}\lf
x\rg={-2x\over x^2-1}\cdot\eeq  Also $m_{\rm D}$ is the Dirac mass
matrix of $\nu_i$'s and $m_{\rm D}^\dag m_{\rm D}$ in \Eref{el}
can be written as follows:
\beq\mD[]^\dag\mD[]=U^{c\dag} d^\dag_{\rm D}d_{\rm D}U^c.
\label{mDD}\eeq \eeqs where $U^c$ are the $3\times3$ unitary
matrix which relates $\sni$ in the $\sni$-basis with the
corresponding in the weak basis. With the help of the seesaw
formula, $\mD[i]$ and $\mrh[i]$ involved in \Eref{el} can be
related to the light-neutrino mass matrix $m_\nu$. Working in the
$\sni$-basis, we have
\beq \label{seesaw} m_\nu= -m_{\rm D}\ d_{N^c}^{-1}\ m_{\rm
D}^{\tr}\>\>\>\mbox{where}\>\> \>d_{N^c}=
\diag\lf\mrh[1],\mrh[2],\mrh[3]\rg \eeq with
$\mrh[1]\leq\mrh[2]\leq\mrh[3]$ real and positive. Based on the
analysis of \cref{nMCI, senoguz}, we find $\bar m_\nu$ via \beq
\bar m_\nu=U_\nu^*\ d_\nu\
U^\dag_\nu\>\>\>\mbox{where}\>\>\>d_{\nu}=
\diag\lf\mn[1],\mn[2],\mn[3]\rg\label{mns1}\eeq with $\mn[1]$,
$\mn[2]$ and $\mn[3]$ being the real and positive light neutrino
mass eigenvalues. These can be found assuming \emph{normal
[inverted] ordered} (NO [IO]) $\mn[i]$'s and using a reference
neutrino mass and the observed \cite{valle, lisi} low energy
neutrino mass-squared differences. Also $U_\nu$ is the PMNS matrix
which is a function of the mixing angles $\th_{ij}$ and the
CP-violating Majorana ($\varphi_1$ and $\varphi_2$) and Dirac
($\delta$) phases. Taking also $\mD[i]$ as input parameters we can
construct the complex symmetric matrix \beq W=-d_{\rm D}^{-1}\bar
m_\nu d_{\rm D}^{-1}\label{Wm}\eeq from which we can extract
$d_{N^c}$ as follows \cite{nMCI, senoguz}: \beq
d_{N^c}^{-2}=U^{c\dag}W W^\dag U^c.\label{WW}\eeq Acting this way
-- see \Sref{num} --, we can determine the elements of $U^c$ and
the $\mrh[i]$'s, compute $m_{\rm D}^{\dag}m_{\rm D}$ through
\Eref{mDD} and finally obtain the $\ve_i$'s via \Eref{el}.

\subsection{Post-Inflationary Requirements}\label{cont1}

The success of our post-inflationary scenario can be judged, if,
in addition to the constraints of \Sref{fhi2}, it is consistent
with the following requirements:

\paragraph{5.4.1}  The bounds on $\mrh[1]$:
\beq\label{kin} {\sf \ftn (a)}\>\>\mrh[1]\gtrsim
10\Trh\>\>\>\mbox{and}\>\>\>{\sf \ftn
(b)}\>\>\msn\geq2\mrh[1].\eeq
The first inequality is applied to avoid any erasure of the
produced $Y_L$ due to $\nu^c_1$ mediated inverse decays and
$\Delta L=1$ scatterings \cite{senoguz}. The second bound ensures
that the decay of $\dphi$ into a pair of $\sni$'s is kinematically
allowed for at least one species of the $\sni$'s.

\paragraph{5.4.2} Constraints from neutrino physics. We take as inputs the
best-fit values \cite{valle} -- see also \cref{lisi} -- on the
neutrino mass-squared differences, $\Delta
m^2_{21}=7.62\cdot10^{-3}~{\rm eV}^2$ and $\Delta m^2_{31}=\lf
2.55\left[-2.43\right]\rg\cdot~10^{-3}~{\rm eV}^2$, on the mixing
angles, $\sin^2\theta_{12}=0.32$,
$\sin^2\theta_{13}=0.0246\left[0.025\right]$, and
$\sin^2\theta_{23}=0.613\left[0.6\right]$ and the Dirac phase
$\delta=0.8\pi\left[-0.03\pi\right]$ for NO [IO] $\mn[i]$'s.
Moreover, the sum of $\mn[i]$'s is bounded from above by the
current data \cite{wmap, plcp}, as follows \beq \mbox{$\sum_i$}
\mn[i]\leq0.28~{\eV}~~\mbox{at 95\% c.l.}\label{mnol}\eeq

\paragraph{5.4.3} The observational results on $Y_B$ \cite{wmap, plcp}
\beq Y_B\simeq\lf8.55\pm0.217\rg\cdot10^{-11}~~\mbox{at 95\%
c.l.}\label{BAUwmap}\eeq

\paragraph{5.4.4} The bounds on $Y_{3/2}$ imposed \cite{kohri} by
successful BBN:
\beq  \label{Ygw} \Yg\lesssim\left\{\bem
%\begin{array}{rl}
%
10^{-14}\hfill \cr
10^{-13}\hfill \cr
10^{-12}\hfill \cr \eem
%\end{array}
\right.\>\>\>\mbox{for}\>\>\>\mgr\simeq\left\{\bem
0.69~{\rm TeV},\hfill \cr
10.6~{\rm TeV},\hfill \cr
13.5~{\rm TeV}.\hfill \cr \eem
%\end{array}
\right.\eeq
Here we consider the conservative case where $\Gr$ decays with a
tiny hadronic branching ratio.

\subsection{Numerical Results}\label{num}

\renewcommand{\arraystretch}{1.3}
\begin{table}[!t]
\bec\begin{tabular}{|c||c|c||c|c|c||c|c|}\hline
{\sc Parameters} &  \multicolumn{7}{c|}{\sc Cases}\\\cline{2-8}
&A&B& C & D& E & F&G\\ \cline{2-8} &\multicolumn{2}{c||}{\sc
Normal} & \multicolumn{3}{|c||}{\sc Almost}&
\multicolumn{2}{|c|}{\sc Inverted}
\\& \multicolumn{2}{c||}{\sc Hierarchy}&\multicolumn{3}{|c||}{\sc Degeneracy}&
\multicolumn{2}{|c|}{\sc Hierarchy}\\ \hline
\multicolumn{8}{|c|}{\sc Low Scale Parameters}\\\hline
$\mn[1]/0.1~\eV$&$0.01$&$0.1$&$0.5$ & $0.7$& $0.7$ & $0.5$&$0.49$\\
$\mn[2]/0.1~\eV$&$0.09$&$0.1$&$0.51$ & $0.7$& $0.7$ & $0.51$&$0.5$\\
$\mn[3]/0.1~\eV$&$0.5$&$0.5$&$0.71$ & $0.86$&$0.5$ &
$0.1$&$0.05$\\\hline
$\sum_i\mn[i]/0.1~\eV$&$0.6$&$0.7$&$1.7$ & $2.3$&$1.9$ &
$1.1$&$1$\\ \hline
$\varphi_1$&$0$&$2\pi/3$&$\pi/2$ & $\pi/2$&$0$ & $-3\pi/4$&$\pi/4$\\
$\varphi_2$&$\pi/2$&$\pi/2$ &$\pi/3$& $2\pi/3$&$-2\pi/3$ &
$5\pi/4$&$-2\pi/3$\\\hline
\multicolumn{8}{|c|}{\sc Leptogenesis-Scale Parameters}\\\hline
$\mD[1]/0.1~\GeV$&$16$&$15$&$9$ & $20$&$7$ & $20$&$5$\\
$\mD[2]/\GeV$&$40$&$8.3$&$10.5$ & $10.3$&$7.5$ & $5.3$&$11.8$\\
$\mD[3]/10~\GeV$&$10$&$10$&$3.56$ & $10$&$10$ & $10$&$4$\\\hline
$\mrh[1]/10^{11}~\GeV$&$12.3$&$2.2$&$0.16$ & $0.58$&$0.11$ & $0.7$&$0.12$\\
$\mrh[2]/10^{12}~\GeV$&$22.2$&$1.8$&$1.8$ & $1.75$&$1$ & $1.6$&$2.2$\\
$\mrh[3]/10^{14}~\GeV$&$25$&$4$&$0.15$ & $0.73$&$0.74$ &
$2.7$&$1.2$\\\hline
\multicolumn{8}{|c|}{\sc Open Decay Channels of the Inflaton,
$\dphi$, Into $\sni$}\\\hline
$\dphi\ \to$&$\wrhn[1]$&$\wrhn[1,2]$& $\wrhn[1,2,3]$&
$\wrhn[1,2]$& $\wrhn[1,2]$ & $\wrhn[1,2]$&$\wrhn[1,2]$\\ \hline
$\GNsn/\Gsn~(\%)$&$3$&$7$& $7$& $6.5$& $2.3$ & $5$&$9.8$\\ \hline
\multicolumn{8}{|c|}{\sc Resulting $B$-Yield }\\\hline
$10^{11}Y_B$&$8.54$&$8.7$&$8.7$ & $8.5$&$8.4$ &
$8.4$&$8.5$\\\hline
\multicolumn{8}{|c|}{\sc Resulting $\Trh$ and $\Gr$-Yield
}\\\hline
$\Trh/10^{8}~\GeV$&$5.9$&$5.9$&$6.3$ & $5.9$&$5.8$ & $5.9$&$6.1$\\
$10^{13}Y_{3/2}$&$1.1$&$1.1$&$1.2$ & $1.13$&$1.11$ &
$1.1$&$1.15$\\\hline
\end{tabular}\ec
\hfill \vchcaption[]{\sl\small  Parameters yielding the correct
BAU for various neutrino mass schemes, $\ksh=1.5,~\lm=10^{-6}$ and
$y=0.5$. Shown also are the branching ratios of the $\dphi$ decay
into $\sni$ with $i=2$ except for the case A where $i=1$. Recall
that these results are independent of the variables $\ld, \ck,
\kx$ and $\ksn$. } \label{tab2}
\end{table}

As shown in \Sref{lept}, \FHI predicts a constant value of $\msn$.
Consequently, $\Trh$ and $Y_B$ -- see \eqs{Trh}{Yb} -- are largely
independent of the precise value of $\ck$ and $\ld$ in the range
of \Eref{res1} -- contrary to the case of FHI \cite{susyhybrid,
rob}. Just for definiteness we specify that throughout this
section we take $\ck=150$ which corresponds to $\ld=0.0034$,
$\ns=0.963$ and $\msn=3\cdot10^{13}~\GeV$. On the other hand,
$\Trh$ and $Y_B$ depend on $\lm$, $y$ and the masses of the
$\sni$'s into which $\dphi$ decays. Throughout our computation we
take $y=0.5$, which is a typical value encountered \cite{fermionM}
into various MSSM settings with large $\tan\beta$, and so the
corresponding decay width via \Eref{Gpq1} is confined to
$\Gysn=0.45~\GeV$. Note that varying $y$ in its plausible
\cite{fermionM} range $(0.4-0.6)$, $\Gysn$ ranges from $0.28$ to
$0.64~\GeV$ causing minor changes to our results.

Following the bottom-up approach described in \Sref{lept2}, we
find the $\mrh[i]$'s by using as inputs the $\mD[i]$'s, a
reference mass of the $\nu_i$'s -- $\mn[1]$ for NO $\mn[i]$'s, or
$\mn[3]$ for IO $\mn[i]$'s --, the two Majorana phases $\varphi_1$
and $\varphi_2$ of the PMNS matrix, and the best-fit values,
mentioned in \Sref{cont1}, for the low energy parameters of
neutrino physics. In our numerical code, we also estimate,
following \cref{running}, the RG evolved values of the latter
parameters at the scale of nTL, $\Lambda_L=\msn$, by considering
the MSSM with $\tan\beta\simeq50$ as an effective theory between
$\Lambda_L$ and the SSB scale, $M_{\rm SUSY}=1.5~\TeV$. We
evaluate the $\mrh[i]$'s at $\Lambda_L$, and we neglect any
possible running of the $\mD[i]$'s and $\mrh[i]$'s. Therefore, we
present their values at $\Lambda_L$.

Fixing $\lm$ at an intermediate value in its allowed region -- see
\Eref{lm} -- $\lm=10^{-6}$ which results, via \Eref{Ghh} in
$\Ghsn=1.3~\GeV$ we can get a first picture for the parameters
which yield $\Yb$ and $\Yg$ compatible with \eqs{BAUwmap}{Ygw},
respectively in \Tref{tab2}. We consider strongly NO (cases A and
B), almost degenerate (cases C, D and E) and strongly IO (cases F
and G) $\mn[i]$'s. In all cases the current limit of \Eref{mnol}
is safely met -- in the case D this limit is almost saturated. We
observe that with NO or IO $\mn[i]$'s, the resulting $\mrh[i]$'s
are also hierarchical. With degenerate $\mn[i]$'s, the resulting
$M_{i\nu}$'s are closer to one another. Consequently, in the
latter case more $\dphi$-decay channels are available, whereas for
the case A only a single decay channel is open. In all other cases
-- even in the case C where the decay channel $\dphi\to
\wrhn[3]\wrhn[3]$ is kinematically permitted --, the dominant
contributions to $\Yb$ arise from $\ve_2$. Therefore, the
branching ratios, which are also presented in \Tref{tab2},
$\GNsn/\Gsn$ with $i=1$ for the case A and $i=2$ for the other
cases are crucial for the calculation $\Yb$ from \Eref{Yb}. We
notice that these ratios introduce a considerable reduction in the
derivation of $\Yb$, given that $\GNsn<\Gysn<\Ghsn$. This
reduction can be eluded if we adopt -- as in \cref{nmH,
susyhybrid,rob} -- the resolution of the $\mu$ problem proposed in
\cref{rsym} since then the decay mode in \Eref{Lh} disappears. In
\Tref{tab2} shown also are the values of $\Trh$, the majority of
which are close to $6\cdot10^8~\GeV$, and the corresponding
$Y_{3/2}$'s, which are consistent with \Eref{Ygw} mostly for
$m_{3/2}\gtrsim11~\TeV$. These large values are in nice agreement
with the ones needed for the solution of the $\mu$ problem of
MSSM, as explained in \Sref{secmu}.

Since we do not consider any particular GUT here, the $\mD[i]$'s
are free parameters. For the sake of comparison, however, we
mention that the simplest realization of a SUSY Left-Right
[Pati-Salam] GUT predicts \cite{rob, nick} $h_{iN}=h_{iE}$
[$\mD[i]=m_{iU}$], where $m_{iU}$ are the masses of the up-type
quarks and we ignore any possible mixing between generations.
Taking into account the SUSY threshold corrections \cite{fermionM}
in the context of MSSM with universal gaugino masses and
$\tan\beta\simeq50$ -- favored by the recent LHC results
\cite{lhc} -- these predictions are translated as follows:
\beq \lf m_{1{\rm D}}^0,m_{2{\rm D}}^0,m_{3{\rm D}}^0 \rg\simeq
\left\{\bem
(0.023,4.9,100)~\GeV &\mbox{for a Left-Right GUT},\hfill \cr
(0.0005,0.24,100)~\GeV &\mbox{for a Pati-Salam GUT}.\hfill\cr \eem
%\end{array}
\right.\eeq
Comparing these values with those listed in \Tref{tab2}, we remark
that our model is not compatible with any GUT-inspired pattern of
large hierarchy between the $\mD[i]$'s, especially in the two
lighter generations, since $\mD[1]\gg\mD[1]^0$ and
$\mD[2]>\mD[2]^0$. On the other hand, in the cases A, B, D, E and
F we are able to place $\mD[3]\simeq m_{3{\rm D}}^0$. This
arrangement can be understand if we take into account that
$\mD[1]$ and $\mD[2]$ separately influences the derivation of
$\mrh[1]$ and $\mrh[2]$ respectively -- see, e.g.,
\cref{nmH,senoguz}. Consequently, the displayed
$\mD[2]\sim10~\GeV$ assists us to obtain the $\ve_2$'s required by
\Eref{BAUwmap} -- note that in the case A $\mD[2]\simeq40~\GeV$
kinematically blocks the channel $\dphi\to \wrhn[2]\wrhn[2]$. On
the other hand, $\mD[1]\gtrsim0.5~\GeV$ is necessitated in order
to obtain the observationally favored $\ve_1$ in the case A and
fulfill \sEref{kin}{a} in the other cases. Note that the phases
$\varphi_1$ and $\varphi_2$ in \Tref{tab2} are selected in each
case, so that the required $\mD[i]$ and $\mrh[i]$, which dominate
the $Y_B$ calculation, and the resulting $\Trh$ are almost
minimized.

%%%%%%%%%%%%%%%%%%%%%%%%%%%%%%%%%%%%%%%%%%%%%%%%%%%%%%%%%%%%%%%%%%%%
\begin{figure}[!t]\vspace*{-.26in}
\begin{center}
\epsfig{file=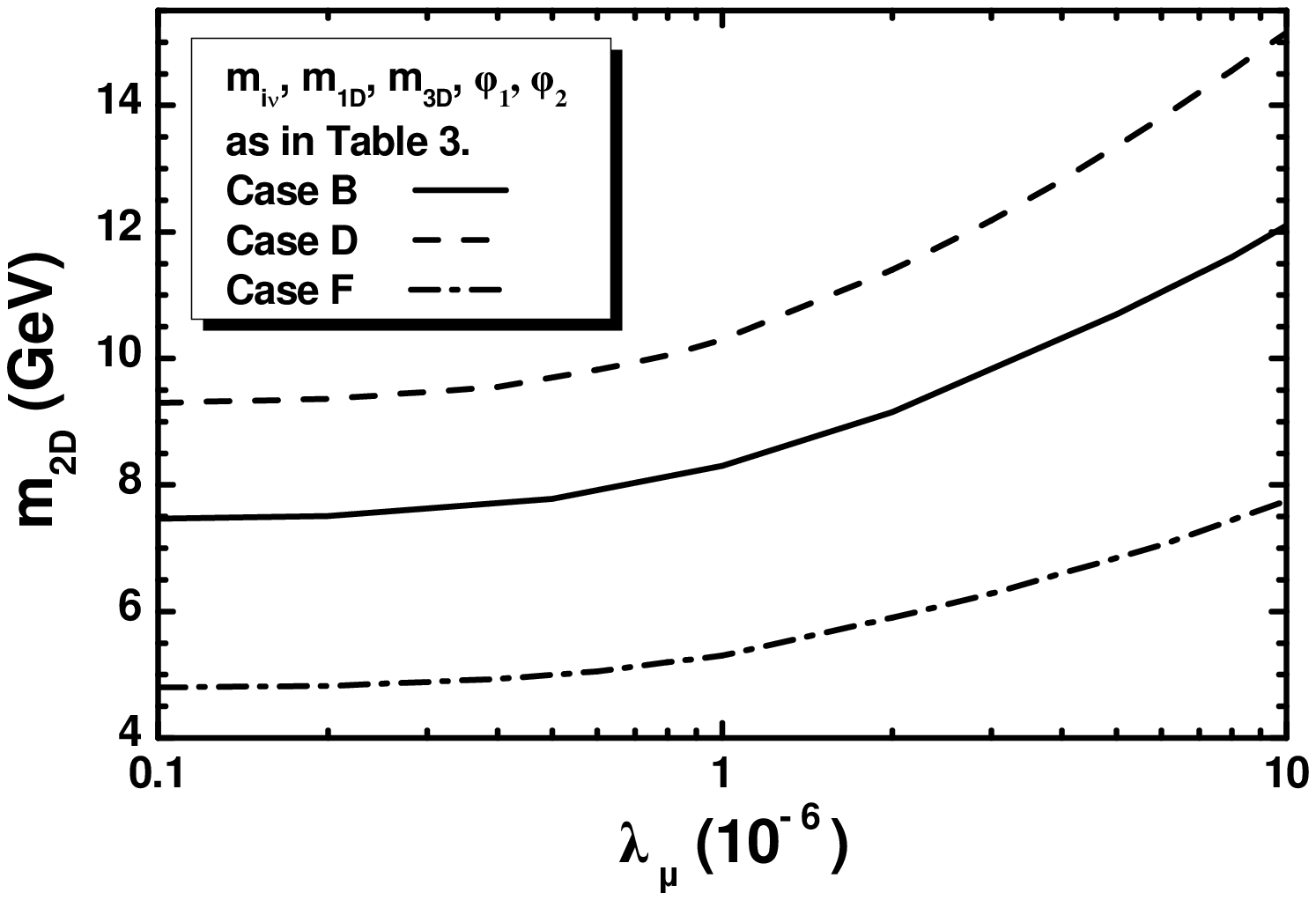,height=3.65in,angle=-90}\ec
\vspace*{-.15in}\hfill \vchcaption[]{\sl \small Contours in the
$\lm-m_{\rm 2D}$ plane yielding the central $Y_B$ in
\Eref{BAUwmap} consistently with the inflationary requirements for
$\ksh=1.5,~y=0.5$ and the values of $m_{i\nu}$, $m_{\rm 1D}$,
$m_{\rm 3D}$, $\varphi_1$, and $\varphi_2$ which correspond to the
cases B (solid line), D (dashed line), and F (dot-dashed line) of
\Tref{tab2}.}\label{kpmD}
\end{figure}
%%%%%%%%%%%%%%%%%%%%%%%%%%%%%%%%%%%%%%%%%

In order to extend the conclusions inferred from \Tref{tab2} to
the case of a variable $\lm$, we can examine how the central value
of $Y_B$ in \Eref{BAUwmap} can be achieved by varying $m_{\rm 2D}$
as a function of $\lm$. The resulting contours in the $\kp-m_{\rm
2D}$ plane are presented in \Fref{kpmD} -- since the range of
$Y_B$ in \Eref{BAUwmap} is very narrow, the $95\%$ c.l. width of
these contours is negligible. The convention adopted for these
lines is also described in the figure. In particular, we use
solid, dashed, or dot-dashed line for $\mn[i]$, $\mD[1]$,
$\mD[3]$, $\varphi_1$, and $\varphi_2$ corresponding to the cases
B, D, or F of \Tref{tab2} respectively. Since increasing $\lm$,
the resulting $\Trh$ is expected to get larger than that shown in
\Tref{tab2} -- see \eqs{Ghh}{Trh} -- we select for the plot in
\Fref{kpmD} one case from every low-energy mass scheme of
$\mn[i]$'s with $\mrh[1]$ large enough, such that \sEref{kin}{a}
is comfortably satisfied for every $\lm$ within the range of
\Eref{lm} with $\ksh=1.5$. This equation sets, actually, the
limits on the contours depicted in \Fref{kpmD}. For
$\lm\gtrsim6\cdot10^{-7}$ we get $\Ghsn>\Gysn$ and so, increasing
$\lm$ the branching fraction in \sEref{Yb}{b} drops and larger
$\mD[2]$'s are required to obtain $Y_B$ compatible with
\Eref{BAUwmap}. On the other hand, for $\lm\lesssim6\cdot10^{-7}$,
$\Gysn$ gets larger than $\Ghsn$ and so, the branching fraction in
\sEref{Yb}{b} remains almost constant and no sizable variation of
$\mD[2]$ is required. At the upper termination points of the
contours, we obtain $\Trh\simeq5\cdot10^9~\GeV$ or
$Y_{\Gr}\simeq9.4\cdot10^{-13}$. The constraint of \Eref{Ygw},
therefore, will cut any possible extension of the curves would be
available for possible larger $\lm$'s. Along the depicted
contours, the resulting $\mrh[2]$'s vary in the range
$(1.4-4)\cdot10^{12}~\GeV$ whereas $\mrh[1]$ and $\mrh[3]$ remain
close to their values presented in the corresponding cases of
\Tref{tab2}.

In conclusion, nTL is a realistic possibility within our model,
thanks to the spontaneously arising couplings in SUGRA, even
without direct couplings of the inflaton to $\sni$'s in $W$.

\section{Conclusions}\label{con}

We investigated a variant of the Starobinsky inflation, which can
be embedded in a moderate extension of MSSM supplemented by three
$\sni$'s and two more superfields, the inflaton and an accompanied
field. Key role in our proposal plays a continuous $R$ symmetry,
which is reduced to the well-known $R$-parity of MSSM, a
$\mathbb{Z}_2$ discrete symmetry and a no-scale-type symmetry
imposed on the \Km. The adopted symmetries have a number of
ensuing consequences: (i) The inflaton appears quadratically in
the super- and \Ka s; (ii) it couples to $\sni$ via SUGRA-induced
interactions ensuring low $\Trh$ and no important contributions to
the one-loop radiative corrections; (iii) the $\mu$ problem of
MSSM can be elegantly resolved provided that a related parameter
in \sup\ is somehow suppressed. The last issue can be naturally
incorporated in various schemes of SSB with relatively large -- of
the order $(10^{4}-10^6)~\GeV$ -- $\mgr$'s which facilitate the
explanation of the recently observed mass of the electroweak Higgs
and the satisfaction of the $\Gr$ constraint.

The next important modification of our set-up compared to other
incarnations -- cf. \cref{linde, eno7,zavalos} -- of the
Starobinsky inflation in SUGRA is the introduction of a variable
scale ($M$) -- besides the existing one in \cref{linde,
eno7,zavalos} -- in the \sup\ and a parameter ($\ck$) in the \Ka\
which was ultimately confined in the range
$76\leq\ck\leq1.5\cdot10^5$. One of these parameters ($M$ and
$\ck$) can be eliminated demanding that the gravitational strength
takes its conventional value at the SUSY vacuum of the theory.
Actually our inflationary model interpolates between the
Starobinsky \cite{R2} and the induced-gravity
\cite{induced,higgsflaton} inflation. Variation of the free model
parameters ($\ld$ and $\ck$) gives us the necessary flexibility in
order to obtain inflation for \sub\ values of the inflaton.
Consequently, our proposal is stable against possible corrections
from higher order terms in the super- and/or \Ka s. Moreover, we
showed that the one-loop radiative corrections remain subdominant
during inflation and the corresponding effective theory is
trustable up to $\mP$.

Despite the addition of the extra parameters, our scheme remains
very predictive since all the possible sets ($\ld,\ck$) which are
compatible with the two inflationary requirements, concerning the
number of the e-foldings and the normalization of the curvature
perturbation, yield almost constant values of $r$ and $\ns$ and a
unique inflaton mass, $\msn$. In particular, we find
$\ns\simeq0.963$, $\as\simeq-0.00068$ and $r\simeq0.0038$, which
are in excellent agreement with the current data, and
$\msn=3\cdot10^{13}~\GeV$. Moreover, the post-inflationary
evolution within our model remains intact from the variation of
the inflationary parameters ($\ld$ and $\ck$). Implementing the
(type I) seesaw mechanism for the generation of the light neutrino
masses, we restricted their Dirac masses, $\mD[i]$, and the masses
of $\sni$'s, $\mrh[i]$, fulfilling a number of requirements, which
originate from the BAU, the (unstable) $\Gr$ abundance and the
neutrino oscillation parameters. Namely, we found
$\mD[1]\geq0.5~\GeV$ and $\mD[2]\simeq10~\GeV$ resulting mostly to
$\mrh[1]\simeq10^{11}~\GeV$ and $\mrh[2]\simeq10^{12}~\GeV$.

As a bottom line, we would like to emphasize that the
Starobinsky-type inflation in no-scale SUGRA can be linked to the
phenomenology of MSSM, even if it is not realized by a matter-like
inflaton as in \cref{eno9}. In our framework, this type of
inflation, driven by a modulus-like field, suggests a resolution
of the $\mu$ problem of MSSM, compatible with large values of
$\mgr$ and it is followed by a robust cosmological scenario --
already applied in many inflationary settings
\cite{nmH,nMCI,susyhybrid,lept,spontaneous,rob} -- ensuring
spontaneous nTL reconcilable with the $\Gr$ constraint and the
neutrino oscillation parameters.

\begin{acknowledgement}

The author would like to cordially thank G. Lazarides,
A.~Mazumdar, Q. Shafi and O. Vives for helpful discussions. This
research was supported by the Generalitat Valenciana under
contract PROMETEOII/2013/017.

\end{acknowledgement}

\def\prdn#1#2#3#4{{\sl Phys. Rev. D }{\bf #1}, no. #4, #3 (#2)}
\def\jcapn#1#2#3#4{{\sl J. Cosmol. Astropart.
Phys. }{\bf #1}, no. #4, #3 (#2)}

\newpage

\rhead[\fancyplain{}{ \bf \thepage}]{\fancyplain{}{\sl Linking
Starobinsky-Type Inflation in no-Scale SUGRA to MSSM}}
\lhead[\fancyplain{}{\sl \leftmark}]{\fancyplain{}{\bf \thepage}}
\cfoot{}

\end{document}